\documentclass[twocolumn]{aastex631}

\renewcommand{\micron}{$\mu$m}
\newcommand{\msun}{M$_\odot$}
\newcommand{\HST}{{\it HST}}
\newcommand{\JWST}{{\it JWST}}

\received{October 20, 2022}
\accepted{November 29, 2022}

\submitjournal{ApJL}

\usepackage{savesym}
\usepackage{amsmath}

\savesymbol{tablenum}
\usepackage{siunitx}
\restoresymbol{SIX}{tablenum}

\let\oldenddeluxetable\enddeluxetable
\let\olddeluxetable\deluxetable
\makeatletter
\renewenvironment{deluxetable}[1]{
\olddeluxetable{[#1]}
\def\pt@format{#1}%
}{\oldenddeluxetable}
\makeatother

\begin{document}

\title{PHANGS--\JWST\ First Results: A statistical view on bubble evolution in NGC~628}

\correspondingauthor{Elizabeth~J.~Watkins}
\email{elizabeth.watkins@uni-heidelberg.de}

\suppressAffiliations

\author[0000-0002-7365-5791]{Elizabeth~J.~Watkins}
\affiliation{Astronomisches Rechen-Institut, Zentrum f\"{u}r Astronomie der Universit\"{a}t Heidelberg, M\"{o}nchhofstra\ss e 12-14, D-69120 Heidelberg, Germany}

\author[0000-0003-0410-4504]{Ashley~T.~Barnes}
\affiliation{Argelander-Institut f\"{u}r Astronomie, Universit\"{a}t Bonn, Auf dem H\"{u}gel 71, 53121, Bonn, Germany}

\author[0000-0001-7448-1749]{Kiana~Henny}
\affiliation{Department of Physics and Astronomy, University of Wyoming, Laramie, WY 82071, USA}

\author[0000-0003-4770-688X]{Hwihyun~Kim}
\affiliation{Gemini Observatory/NSF’s NOIRLab, 950 N. Cherry Avenue, Tucson, AZ, USA}

\author[0000-0001-6551-3091]{Kathryn~Kreckel}
\affiliation{Astronomisches Rechen-Institut, Zentrum f\"{u}r Astronomie der Universit\"{a}t Heidelberg, M\"{o}nchhofstra\ss e 12-14, D-69120 Heidelberg, Germany}

\author[0000-0002-6118-4048]{Sharon E. Meidt}
\affiliation{Sterrenkundig Observatorium, Universiteit Gent, Krijgslaan 281 S9, B-9000 Gent, Belgium}

\author[0000-0002-0560-3172]{Ralf S.\ Klessen}
\affiliation{Universit\"{a}t Heidelberg, Zentrum f\"{u}r Astronomie, Institut f\"{u}r Theoretische Astrophysik, Albert-Ueberle-Stra{\ss}e 2, D-69120 Heidelberg, Germany}
\affiliation{Universit\"{a}t Heidelberg, Interdisziplin\"{a}res Zentrum f\"{u}r Wissenschaftliches Rechnen, Im Neuenheimer Feld 205, D-69120 Heidelberg, Germany}

\author[0000-0001-6708-1317]{Simon~C.~O.~Glover}
\affiliation{Universit\"{a}t Heidelberg, Zentrum f\"{u}r Astronomie, Institut f\"{u}r Theoretische Astrophysik, Albert-Ueberle-Stra{\ss}e 2, D-69120 Heidelberg, Germany}

\author[0000-0002-0012-2142]{Thomas G. Williams}
\affiliation{Sub-department of Astrophysics, Department of Physics, University of Oxford, Keble Road, Oxford OX1 3RH, UK}
\affiliation{Max-Planck-Institut f\"ur Astronomie, K\"onigstuhl 17, D-69117 Heidelberg, Germany}

\author[0000-0002-9642-7193]{Benjamin~W.~Keller}
\affiliation{Department of Physics and Materials Science, University of Memphis, 3720 Alumni Avenue, Memphis, TN 38152, USA}

\author[0000-0002-2545-1700]{Adam~K.~Leroy}
\affiliation{Department of Astronomy, The Ohio State University, 140 West 18th Avenue, Columbus, Ohio 43210, USA}
\affiliation{Center for Cosmology and Astroparticle Physics, 191 West Woodruff Avenue, Columbus, OH 43210, USA}

\author[0000-0002-5204-2259]{Erik~Rosolowsky}
\affiliation{Department of Physics, University of Alberta, Edmonton, Alberta, T6G 2E1, Canada}

\author[0000-0003-0946-6176]{Janice C. Lee}
\affiliation{Gemini Observatory/NSF’s NOIRLab, 950 N. Cherry Avenue, Tucson, AZ, USA}
\affiliation{Steward Observatory, University of Arizona, 933 N Cherry Ave, Tucson, AZ 85721, USA}

\author[0000-0002-5259-2314]{Gagandeep S. Anand}
\affiliation{Space Telescope Science Institute, 3700 San Martin Drive, Baltimore, MD 21218, USA}

\author[0000-0002-2545-5752]{Francesco Belfiore}
\affiliation{INAF -- Osservatorio Astrofisico di Arcetri, Largo E. Fermi 5, I-50157, Firenze, Italy}

\author[0000-0003-0166-9745]{Frank Bigiel}
\affiliation{Argelander-Institut f\"{u}r Astronomie, Universit\"{a}t Bonn, Auf dem H\"{u}gel 71, 53121, Bonn, Germany}

\author[0000-0003-4218-3944]{Guillermo A. Blanc}
\affiliation{The Observatories of the Carnegie Institution for Science, 813 Santa Barbara St., Pasadena, CA, USA}
\affiliation{Departamento de Astronom\'{i}a, Universidad de Chile, Camino del Observatorio 1515, Las Condes, Santiago, Chile}

\author[0000-0003-0946-6176]{Médéric~Boquien}
\affiliation{Centro de Astronomía (CITEVA), Universidad de Antofagasta, Avenida Angamos 601, Antofagasta, Chile}

\author[0000-0001-5301-1326]{Yixian Cao}
\affiliation{Max-Planck-Institut f\"ur Extraterrestrische Physik (MPE), Giessenbachstr. 1, D-85748 Garching, Germany}

\author[0000-0003-0085-4623]{Rupali~Chandar}
\affiliation{Ritter Astrophysical Research Center, University of Toledo, Toledo, OH 43606, USA}

\author[0000-0002-5993-6685]{Ness Mayker Chen}
\affiliation{Department of Astronomy, The Ohio State University, 140 West 18th Avenue, Columbus, Ohio 43210, USA}
\affiliation{Center for Cosmology and Astroparticle Physics, 191 West Woodruff Avenue, Columbus, OH 43210, USA}

\author[0000-0002-5635-5180]{M\'elanie Chevance}
\affiliation{Universit\"{a}t Heidelberg, Zentrum f\"{u}r Astronomie, Institut f\"{u}r Theoretische Astrophysik, Albert-Ueberle-Stra{\ss}e 2, D-69120 Heidelberg, Germany}
\affiliation{Cosmic Origins Of Life (COOL) Research DAO, coolresearch.io}

\author[0000-0002-8549-4083]{Enrico Congiu}
\affiliation{Departamento de Astronom\'{i}a, Universidad de Chile, Camino del Observatorio 1515, Las Condes, Santiago, Chile}

\author[0000-0002-5782-9093]{Daniel~A.~Dale}
\affiliation{Department of Physics and Astronomy, University of Wyoming, Laramie, WY 82071, USA}

\author[0000-0003-1943-723X]{Sinan Deger}
\affiliation{The Oskar Klein Centre for Cosmoparticle Physics, Department of Physics, Stockholm University, AlbaNova, Stockholm, SE-106 91, Sweden}

\author[0000-0002-4755-118X]{Oleg V. Egorov}
\affiliation{Astronomisches Rechen-Institut, Zentrum f\"{u}r Astronomie der Universit\"{a}t Heidelberg, M\"{o}nchhofstra\ss e 12-14, D-69120 Heidelberg, Germany}

\author[0000-0002-6155-7166]{Eric Emsellem}
\affiliation{European Southern Observatory, Karl-Schwarzschild-Stra{\ss}e 2, 85748 Garching, Germany}
\affiliation{Univ Lyon, Univ Lyon1, ENS de Lyon, CNRS, Centre de Recherche Astrophysique de Lyon UMR5574, F-69230 Saint-Genis-Laval France}

\author[0000-0001-5310-467X]{Christopher M. Faesi}
\affiliation{University of Connecticut, Department of Physics, 196A  Auditorium Road, Unit 3046, Storrs, CT, 06269}

\author[0000-0002-3247-5321]{Kathryn~Grasha}
\affiliation{Research School of Astronomy and Astrophysics, Australian National University, Canberra, ACT 2611, Australia} 
\affiliation{ARC Centre of Excellence for All Sky Astrophysics in 3 Dimensions (ASTRO 3D), Australia}   

\author[0000-0002-9768-0246]{Brent Groves}
\affiliation{International Centre for Radio Astronomy Research, University of Western Australia, 7 Fairway, Crawley, 6009 WA, Australia}

\author[0000-0002-8806-6308]{Hamid Hassani}
\affiliation{Department of Physics, University of Alberta, Edmonton, Alberta, T6G 2E1, Canada}

\author[0000-0001-9656-7682]{Jonathan~D.~Henshaw}
\affiliation{Astrophysics Research Institute, Liverpool John Moores University, 146 Brownlow Hill, Liverpool L3 5RF, UK}
\affiliation{Max-Planck-Institut f\"ur Astronomie, K\"onigstuhl 17, D-69117 Heidelberg, Germany}

\author{Cinthya Herrera}
\affiliation{IRAM, 300 rue de la Piscine, F-38406 Saint Martin d’H\`eres, France}

\author{Annie Hughes}
\affiliation{CNRS, IRAP, 9 Av. du Colonel Roche, BP 44346, F-31028 Toulouse cedex 4, France}
\affiliation{Universit\'{e} de Toulouse, UPS-OMP, IRAP, F-31028 Toulouse cedex 4, France}

\author[0000-0002-4232-0200]{Sarah Jeffreson}
\affiliation{Center for Astrophysics, Harvard \& Smithsonian, 60 Garden Street, Cambridge MA, United States}

\author[0000-0002-9165-8080]{Mar\'ia J. Jim\'enez-Donaire}
\affiliation{Observatorio Astron\'omico Nacional, Alfonso XII 3, 28014, Madrid, Spain}
\affiliation{Centro de Desarrollos Tecnológicos, Observatorio de Yebes (IGN), 19141 Yebes, Guadalajara, Spain}

\author[0000-0001-9605-780X]{Eric W. Koch}
\affiliation{Center for Astrophysics, Harvard \& Smithsonian, 60 Garden Street, Cambridge MA, United States}

\author[0000-0002-8804-0212]{J.~M.~Diederik~Kruijssen}
\affiliation{Cosmic Origins Of Life (COOL) Research DAO, coolresearch.io}

\author[0000-0003-3917-6460]{Kirsten L. Larson}
\affiliation{AURA for the European Space Agency (ESA), Space Telescope Science Institute, 3700 San Martin Drive, Baltimore, MD 21218, USA}

\author[0000-0001-9773-7479]{Daizhong Liu}
\affiliation{Max-Planck-Institut f\"ur Extraterrestrische Physik (MPE), Giessenbachstr. 1, D-85748 Garching, Germany}

\author[0000-0002-1790-3148]{Laura A. Lopez}
\affiliation{Department of Astronomy, The Ohio State University, 140 West 18th Avenue, Columbus, Ohio 43210, USA}
\affiliation{Center for Cosmology and Astroparticle Physics, 191 West Woodruff Avenue, Columbus, OH 43210, USA}
\affiliation{Flatiron Institute, Center for Computational Astrophysics, NY 10010, USA}

\author[0000-0002-0873-5744]{Ismael Pessa}
\affiliation{Max-Planck-Institut f\"ur Astronomie, K\"onigstuhl 17, D-69117 Heidelberg, Germany}
\affiliation{Leibniz-Institut f\"{u}r Astrophysik Potsdam (AIP), An der Sternwarte 16, 14482 Potsdam, Germany}

\author[0000-0003-3061-6546]{J\'er\^ome Pety}
\affiliation{IRAM, 300 rue de la Piscine, 38400 Saint Martin d'H\`eres, France}
\affiliation{LERMA, Observatoire de Paris, PSL Research University, CNRS, Sorbonne Universit\'es, 75014 Paris}

\author[0000-0002-0472-1011]{Miguel~Querejeta}
\affiliation{Observatorio Astron\'omico Nacional, Alfonso XII 3, 28014, Madrid, Spain}

\author[0000-0002-2501-9328]{Toshiki Saito}
\affiliation{National Astronomical Observatory of Japan, 2-21-1 Osawa, Mitaka, Tokyo, 181-8588, Japan}

\author[0000-0002-4378-8534]{Karin Sandstrom}
\affiliation{Center for Astrophysics and Space Sciences, Department of Physics, University of California, San Diego\\9500 Gilman Drive, La Jolla, CA 92093, USA}

\author[0000-0003-2707-4678]{Fabian Scheuermann}
\affiliation{Astronomisches Rechen-Institut, Zentrum f\"{u}r Astronomie der Universit\"{a}t Heidelberg, M\"{o}nchhofstra\ss e 12-14, D-69120 Heidelberg, Germany}

\author[0000-0002-3933-7677]{Eva Schinnerer}
\affiliation{Max-Planck-Institut f\"ur Astronomie, K\"onigstuhl 17, D-69117 Heidelberg, Germany}

\author[0000-0001-6113-6241]{Mattia C. Sormani}
\affiliation{Universit\"{a}t Heidelberg, Zentrum f\"{u}r Astronomie, Institut f\"{u}r Theoretische Astrophysik, Albert-Ueberle-Stra{\ss}e 2, D-69120 Heidelberg, Germany}

\author[0000-0002-9333-387X]{Sophia K. Stuber}
\affiliation{Max-Planck-Institut f\"ur Astronomie, K\"onigstuhl 17, D-69117 Heidelberg, Germany}

\author[0000-0002-8528-7340]{David A. Thilker}
\affiliation{Department of Physics and Astronomy, The Johns Hopkins University, Baltimore, MD 21218, USA}

\author[0000-0003-1242-505X]{Antonio Usero}
\affiliation{Observatorio Astron\'omico Nacional, Alfonso XII 3, 28014, Madrid, Spain}

\author[0000-0002-3784-7032]{Bradley C. Whitmore}
\affiliation{Space Telescope Science Institute, 3700 San Martin Drive, Baltimore, MD, USA}

\begin{abstract}

The first \JWST\ observations of nearby galaxies have unveiled a rich population of bubbles that trace the stellar feedback mechanisms responsible for their creation. Studying these bubbles therefore allows us to chart the interaction between stellar feedback and the interstellar medium, and the larger galactic flows needed to regulate star formation processes globally. We present the first catalog of bubbles in NGC~628, visually identified using MIRI F770W PHANGS--\JWST\ observations, and use them to statistically evaluate bubble characteristics. We classify 1694 structures as bubbles with radii between 6--552~pc. Of these, 31\% contain at least one smaller bubble at their edge, indicating that previous generations of star formation have a local impact on where new stars form. On large scales, most bubbles lie near a spiral arm, and their radii increase downstream compared to upstream. Furthermore, bubbles are elongated in a similar direction to the spiral arm ridge-line.
These azimuthal trends demonstrate that star formation is intimately connected to the spiral arm passage. Finally, the bubble size distribution follows a power-law of index $p=-2.2\pm0.1$, which is slightly shallower than the theoretical value by 1--3.5$\sigma$ that did not include bubble mergers. The fraction of bubbles identified within the shells of larger bubbles suggests that bubble merging is a common process. Our analysis therefore allows us to quantify the number of star-forming regions that are influenced by an earlier generation, and the role feedback processes have in setting the global star formation rate. With the full PHANGS--\JWST\ sample, we can do this for more galaxies.

\end{abstract}


\section{Introduction} \label{sec:intro}

\begin{figure*}
    \centering
    \includegraphics[width = \textwidth]{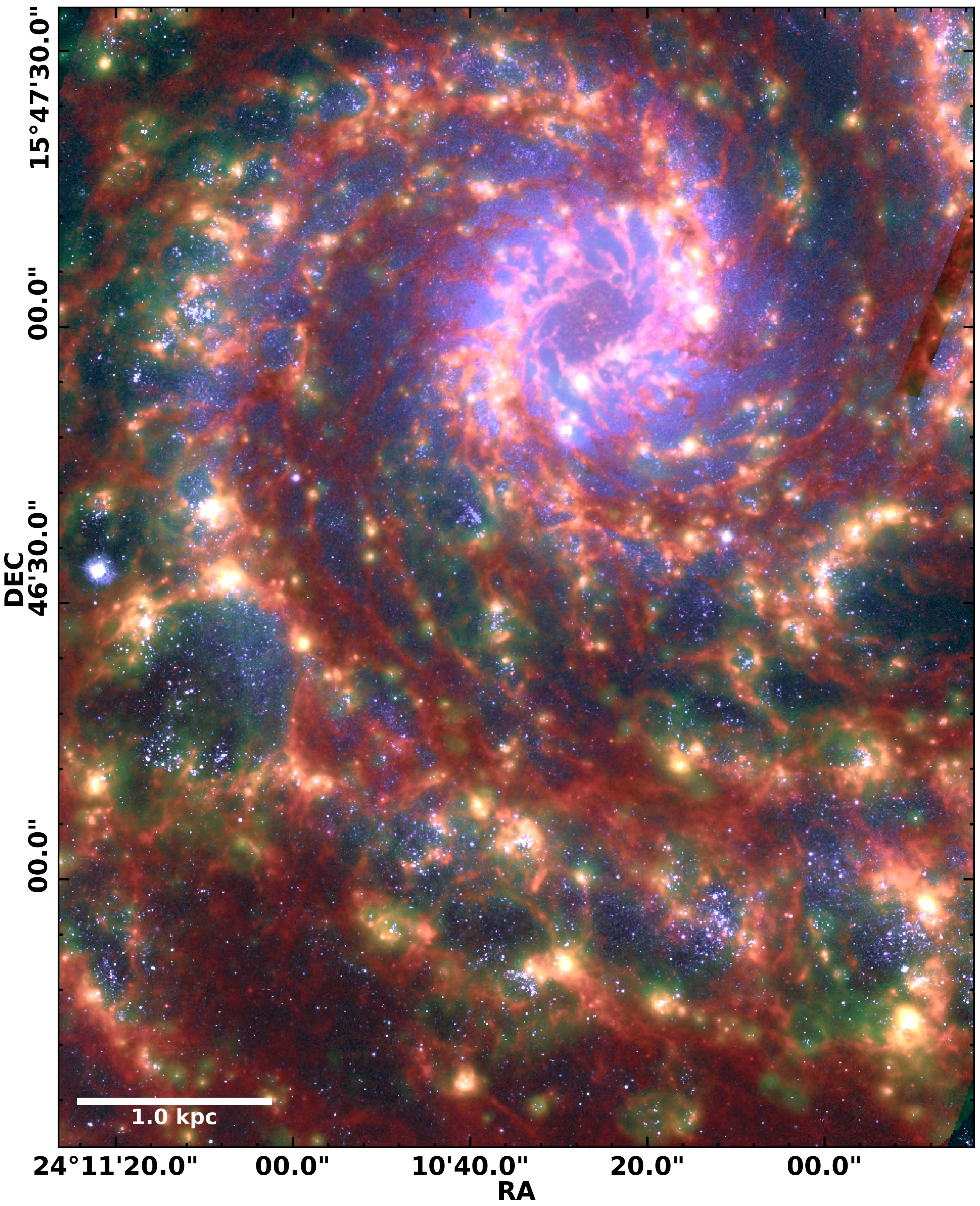}
    \caption{{\bf Multi-scale bubble population in NGC~628 revealed with \JWST\ observations}. RGB illustration towards the center of NGC~628 highlighting bubble features and the bands used in their identification. Red traces MIRI F770W observations, green traces MUSE H$\alpha$ observations and blue traces \textit{B}-band \HST\ observations. White scalebar indicates physical size equal to 1~kpc.}
    \label{fig:bub_rgb}
\end{figure*}

\noindent A ubiquitous feature of new mid-infrared (MIR) James Webb Space Telescope (\JWST) observations of nearby galaxies are the pc to kpc scale holes that riddle the interstellar medium (ISM) (e.g., Figure \ref{fig:bub_rgb}). 
Many of these holes are carved out by stellar feedback processes, and so trace H\textsc{ii} regions, bubbles and superbubbles, although some holes are dynamically driven regions without a powering source, or are remnants of older, sheared bubbles \citep{palous_evolution_1990,kim_three-dimensional_2002,dobbs_exciting_2013}. Most of the smallest holes (hereafter bubbles) with sizes of $\sim$6~pc are driven by pre-supernova (pre-SN) feedback and photoionization from individual high-mass stars and low-mass stellar OB associations \citep{lim_surveying_2020}, though some could be young ($<10^4$~yr) SN remnants. Larger bubbles (up to $\sim$30~pc) are dominated by winds and SNe connected to OB associations \citep{ostriker_astrophysical_1988}. The largest features, superbubbles, represent the combined impact of large, or multiple OB associations that can drive bubbles to radii of $\sim$1000~pc in 40~Myr \citep{nath_size_2020}.

Considering bubbles are such a dominant feature for a range of sizes spanning over two orders of magnitudes (see Figure \ref{fig:bub_rgb}), understanding what creates them, and their evolution, is crucial to understand  how they impact the ISM, their role in driving the galactic-scale star formation cycle, and the processes that result in low star formation efficiencies compared to free-fall timescales \citep{hopkins_galaxies_2014,federrath_inefficient_2015,grudic_nature_2019,keller_empirically_2022}. 
For example, the stellar populations that drive bubbles also contribute to the large scale injection of turbulence that redistributes energy and matter, limiting the large-scale collapse of gas into stars \citep{ostriker_regulation_2010, faucher-giguere_feedback-regulated_2013, krumholz_is_2016, fisher_testing_2019} and leading to galactic scale outflows \citep{fielding_clustered_2018, smith_efficient_2021}. The overall size distribution of bubbles, in particular, provides insight into how clustered star formation is within galaxies, and the pressure balance between different phases of the ISM \citep{nath_size_2020}.

In the Milky Way, we can reach the resolution needed to study and catalog bubbles at the earliest phases of their formation and evolution \citep{simpson_milky_2012, beaumont_milky_2014, anderson_wise_2014, jayasinghe_milky_2019, olivier_evolution_2021}.
However, line of sight confusion within the Milky Way requires complex kinematic analysis \citep{reynolds_optical_1979,ehlerova_origin_1996, ochsendorf_nested_2015,tsujimoto_new_2021-1, zucker_star_2022}, whereas extragalactic observations provide the necessary context to easily connect their shells -- and any co-spatial information such as powering sources -- to larger scale features (\citealt{bagetakos_fine-scale_2011}, Watkins et al., subm.).
Yet, replicating Milky Way studies in extragalactic environments has proven challenging due to the high resolution ($\sim10$~pc) required in atomic and molecular gas to resolve their edges, and in optical emission to resolve the hot ionized gas in their interiors.

With the high ($\sim$10~pc) resolution \JWST\ can reach tracing dust emission within nearby galaxies, we can finally bridge the scale gap between extragalactic and galactic studies. 
Already, \JWST\ observations are revolutionizing our understanding of stellar feedback and the star formation cycle in nearby galaxies over a large range of evolutionary stages. MIR bands, such as MIRI F770W, trace hot dust heated by young stars and polycyclic aromatic hydrocarbons (PAHs), where PAHs are vibrationally excited in the presence of starlight \citep{SANDSTROM1_PHANGSJWST}, especially when illuminated by UV photons. Therefore MIR observations allow us to identify new, young embedded clusters obscured at optical wavelengths \citep{RODRIGUEZ_PHANGSJWST}, large-scale filamentary structures containing dense, cold gas expected to host future star formation \citep{THILKER_PHANGSJWST}, and hot dust emission shining in the presence of UV radiation emitted by OB stars \citep{LEROY1_PHANGSJWST}.  Piecing these results together provides the observations needed to trace recent star formation histories within these galaxies \citep{KIM_PHANGSJWST}.
In this letter we provide a crucial piece needed to understand the star formation picture -- connecting together small and large scale feedback processes -- by cataloging the feedback-driven bubbles in NGC~628 using PHANGS--\JWST.

\section{Observations} \label{sec:obs}

\subsection{NGC~628}
\noindent NGC~628 (also known as M74 or `the Phantom Galaxy') is a nearly face-on spiral galaxy at a distance of 9.84$\pm$0.63~Mpc \citep{mcquinn_accurate_2017,anand_distances_2021}. Its star formation history and multi-phase ISM have been investigated extensively in previous works \citep{sanchez_ppak_2011, grasha_spatial_2015, kreckel_50_2018}, and it has co-spatial PHANGS (Physics at High Angular resolution in
Nearby GalaxieS) observations -- PHANGS--ALMA \citep{leroy_phangs-alma_2021}, PHANGS--MUSE \citep{emsellem_phangs-muse_2022-1} and PHANGS--\HST\ \citep{lee_phangs-hst_2022} -- providing a more comprehensive view of the star formation cycle within this galaxy.

\subsection{\JWST\ observations}
\noindent NGC~628 is one of nineteen galaxies being observed as part of the PHANGS\footnote{ \url{http://www.phangs.org
}}--\JWST\ cycle 1 treasury program using NIRCam's F200W,
F300M, F335M and F360M, and MIRI's F770W, F1000W,
F1130W and F2100W broadband filters (project ID 02107; see \citealt{LEE_PHANGSJWST}). It was observed on July 17, 2022 and the data have been reduced using the publicly available reduction pipeline along with additional reduction tools developed and outlined in \cite{LEE_PHANGSJWST}. These observations cover the central $\ang{;3.8;}\times\ang{;2.2;}$ (11 kpc $\times$ 6.3 kpc) of NGC~628 and for the F770W observations, they have a sensitivity of 0.11~MJy~sr$^{-1}$ and an angular resolution of $\sim\ang{;;0.25}$, which achieves a physical resolution of 12~pc \citep{LEE_PHANGSJWST}.

\subsection{Ancillary data}

\noindent NGC~628 was observed in five bands by \HST\ (WFC3/F275W, WFC3/F336W, ACS/F435W, ACS/F555W, ACS/F814W) as part of the LEGUS survey \citep{calzetti_legacy_2015}, and reduced using the PHANGS--\HST\ data pipeline \citep{lee_phangs-hst_2022}. It was also observed in H$\alpha$ (F658N: Proposal 10402). In this letter, we use the \textit{B}-band observations ($\sim\ang{;;0.08}\sim$4~pc) to visually identify younger sources to help separate the nested bubble population revealed in the \JWST\ observations.

Continuum subtracted H$\alpha$ imaging is taken from the PHANGS--MUSE survey \citep{kreckel_50_2018, emsellem_phangs-muse_2022-1}, an optical IFU survey targeting nineteen nearby galaxies (the same as PHANGS--\JWST) at $\ang{;;0.9}$ ($\sim$50~pc) resolution. This survey traces local ionized gas properties and metallicities, which can be used to constrain  the local star formation history and gas mixing scales. AstroSat (Hassani et al.\ in prep.) far-UV observations trace the ionized emission from older stellar populations at resolutions of $\sim\ang{;;1.5}$, and \textit{Spitzer} observations of 8~\micron\ emission \citep{dale_spitzer_2009} at $\sim\ang{;;2}$ allow us to compare the previous MIR observations to \JWST. We also take into account H\textsc{i} data from the THINGS survey at resolutions of $\sim\ang{;;6}$ \citep{walter_things_2008} and PHANGS--ALMA $^{12}$CO (J=2--1) \citep{leroy_phangs-alma_2021} (at $\sim\ang{;;1}$)  to investigate bubble shells in different bands.

\begin{figure*}
    \centering
        \includegraphics[width=\textwidth]{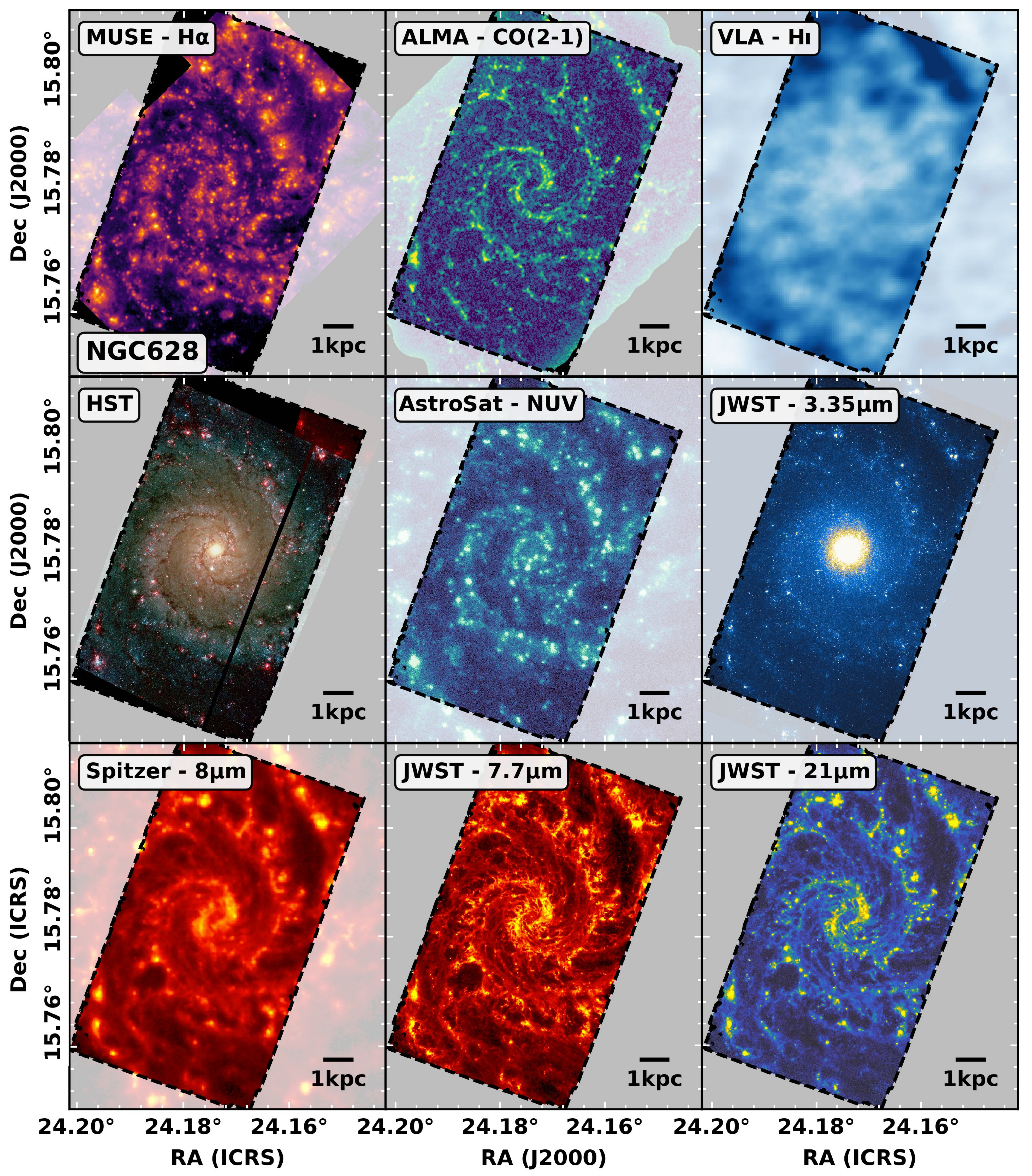}
        \vspace{-7mm}
    \caption{{\bf Multi-wavelength observations covering the nearby star-forming galaxy NGC~628.} In the top row of panels from left to right, we show the H$\alpha$ map from PHANGS--MUSE  \citep{emsellem_phangs-muse_2022-1}, CO\,(2-1) peak intensity from PHANGS--ALMA, \citep{leroy_phangs-alma_2021}, and the VLA H\textsc{i} moment 0 map from the THINGS survey \citep{walter_things_2008}. In the middle row, left panel, we show an image produced from the {\it B} (blue), {\it V} (green) and {\it I} (red) band filters from \HST, and the continuum subtracted \HST--H$\alpha$ (658\,nm; red, see \citealt{lee_phangs-hst_2022}). In the middle center and right panels, we show the  UV emission map from AstroSat (Hassani et al. in prep) and NIRCam F335M emission from \JWST\ \citep{LEE_PHANGSJWST}. On the bottom row, from left to right, we show the 8\,\micron\ map from {\it Spitzer} \citep{dale_spitzer_2009}, the F770W map and the F2100W map from MIRI \JWST\ \citep{LEE_PHANGSJWST}. All panels have been overlaid with a dashed contour showing the coverage of the \JWST\ observations.}
    \label{fig:multiwave}
\end{figure*}

%%%%%%%%%%%%%%%%%%%%%%%%%%%%%%%%%%%%%%%%%%%%%%
\section{A panchromatic picture of bubble features in NGC~628} 
\label{sec_bubsample}

\noindent In Figure \ref{fig:multiwave}, we present a panchromatic picture of NGC~628. Bubble features, such as shells, stellar sources and hot emission enclosed by the bubbles, are detectable throughout the electromagnetic spectrum (from X-ray to radio), and especially pronounced at MIR bands, which is why MIR broadband filters have been used extensively to study bubbles in the Milky Way \citep{churchwell_bubbling_2006,anderson_wise_2014,jayasinghe_milky_2019}. 

For our panchromatic view, we chose bands that highlight the full range of different bubble features. For example, MIRI F770W and F1130W have similar emission features, but we only show F770W maps as they are at a higher resolution. Altogether, we consider AstroSat far-UV, \HST\ \textit{B}, \textit{V} and \textit{I} bands, and narrow band H$\alpha$, MUSE H$\alpha$, NIRCam F335M, MIRI F770W and F2100W, {\it Spitzer} 8~\micron,  ALMA $^{12}$CO (J=2--1) and THINGS H\textsc{i} moment 0 maps. For the NIRCam bands, we only show the F335M band since it contains both stellar sources and some weak PAH emission, potentially allowing us to see both the bubble shells and sources within them.

In F2100W emission, MUSE H$\alpha$ and AstroSat far-UV, we can see concentrated ionized emission, and some hints of ionized shells in MUSE H$\alpha$, but it is difficult to define their boundaries due to confusion with the more diffuse ionized gas \citep{belfiore_tale_2022}. We also note that the vast majority are not resolved in the $\sim$\,50\,pc resolution MUSE observations (see e.g.,\ \citealp{santoro_phangsmuse_2022, barnes_comparing_2021}).

In \HST\ and \JWST\ NIRCam imaging, we see bright clustered sources along the arms, though \HST\ shows them more clearly toward the center of the galaxy. 
While there is PAH emission in the NIRCam F335M map, it is faint compared to the MIRI F770W emission, which is expected given the relative intrinsic strength of the two features.

Molecular gas offers a clear constraint on bubble shells, particularly in the case of the larger ($>50$~pc) bubbles. The high spectral resolution achievable with molecular line observations also offers dynamical information (e.g.,\ CO at 2.5~km~s$^{-1}$; \citealp{leroy_phangs-alma_2021}), providing kinematic confirmation of coherent bubbles structures and constraining properties of the powering sources (Watkins et al. subm.). However, CO provides a limited sample of bubble structures. They must be large enough to be resolved by ALMA ($\sim$\,50\,pc resolution), and they also need to be young enough ($<$8~Myr) to still contain molecular gas \citep{kim_star_2021}. In Watkins et al.\ (subm.), we only find 12 unbroken bubbles in NGC~628 in CO due to these constraints.
H\textsc{i} emission offers an alternative, as can be seen in the THINGS H\textsc{i} map \citep{bagetakos_fine-scale_2011,krause_26al_2015-1} but at 200~pc resolution, it resolves only the largest bubble features \citep{BARNES_PHANGSJWST}.

While lacking the kinematic confirmation, broadband dust (PAH) emission at MIR probes the three gas phases
\citep{LEROY1_PHANGSJWST}. In addition, stellar populations found inside bubbles produce UV photons in abundance \citep{peeters_polycyclic_2004}. As the underlying dust and stellar continuum contributions are relatively weak at $\sim8$~\micron\ \citep{marble_aromatic_2010} and there is increased gas (and therefore dust) column densities in swept-up shells, bubble features stand out against the background. This makes them highly visible at a glance in MIRI F770W observations in Figure \ref{fig:bub_rgb} and Figure \ref{fig:multiwave} at high (12~pc) resolution. While we can still see some larger bubble features towards the top of the image and in the lower left in the {\it Spitzer} 8~\micron\ map, it lacks the resolution to probe smaller bubble features \citep{dale_spitzer_2009}, limiting the number of bubbles we can see.

The high resolution MIRI F770W shows that two physically different structures exist across the galaxy (also see Figure \ref{fig:bub_rgb} and Figure \ref{fig:bub_loc}, and Figure \ref{fig:close-up} in Appendix \ref{sec:close-up}): 
\begin{itemize}
\item[i)] Spherical or elliptical `bubbles' that are closely linked downstream from a given arm (convex side) and are driven by feedback;
\item[ii)] Very elongated elliptical `holes' that are situated between the arms, towards the upstream part of a given arm (concave side) and are old sheared bubbles, or dynamically created holes. 
\end{itemize}

Dynamically, NGC~628 is rotating clockwise and the arms are trailing. Hence, inside co-rotation (4.4$\pm$2.0~kpc; \citealt{williams_applying_2021}), the fact that the disk material moves faster than the spiral puts recent star formation on the downstream side of the arms, with age increasing away from the arm (discussed further in Section \ref{sec:results}).
In this picture, the feedback from the newly-formed star clusters preferentially flows into the bubbles that are already inflated by previous star formation associated with this arm. 

\begin{table*}[]
\centering
% \resizebox{\textwidth}{!}{
\begin{tabular}{rrrrrrrrrr}
Bubble  & RA & Dec & Semi-major  & Semi-minor  & Average     & PA        & Arm & Distance to & Galactocentric  \\
ID      &    &     & axis (pc)   & axis (pc)   & radius (pc) & ($\circ$) &     & arm (pc)    & radius (kpc)    \\
\hline
1 & $\ang{24;08;55.9}$ & $\ang{15;48;7.9}$ & $86$ & $86$ & $86$ & $0$ & $3$ & $145$ & $5.22$\\
2 & $\ang{24;08;56.7}$ & $\ang{15;48;2.9}$ & $24$ & $24$ & $24$ & $0$ & $3$ & $127$ & $5.05$\\
3 & $\ang{24;08;57.0}$ & $\ang{15;48;9.0}$ & $42$ & $42$ & $42$ & $0$ & $3$ & $138$ & $5.21$\\
4 & $\ang{24;08;58.0}$ & $\ang{15;47;44.9}$ & $25$ & $25$ & $25$ & $0$ & $3$ & $101$ & $4.54$\\
5 & $\ang{24;08;58.2}$ & $\ang{15;48;2.6}$ & $24$ & $24$ & $24$ & $0$ & $3$ & $114$ & $4.98$\\
6 & $\ang{24;08;58.7}$ & $\ang{15;47;41.2}$ & $33$ & $33$ & $33$ & $0$ & $3$ & $95$ & $4.44$\\
7 & $\ang{24;08;58.8}$ & $\ang{15;47;50.6}$ & $32$ & $31$ & $32$ & $0$ & $3$ & $93$ & $4.64$\\
8 & $\ang{24;08;58.9}$ & $\ang{15;47;39.8}$ & $40$ & $40$ & $40$ & $0$ & $3$ & $93$ & $4.40$\\
9 & $\ang{24;08;58.9}$ & $\ang{15;48;9.0}$ & $31$ & $31$ & $31$ & $0$ & $3$ & $122$ & $5.14$\\
10 & $\ang{24;08;59.3}$ & $\ang{15;48;6.1}$ & $209$ & $171$ & $196$ & $0$ & $3$ & $112$ & $5.04$\\
\end{tabular}
% }
    \caption{Sample of 10 bubbles ordered by RA and their defining properties. Distance used to calculate physical lengths is 9.84$\pm$0.63~Mpc. See Section \ref{sec:cat-des} for more information about uncertainties.}
    \label{tab:bub_props}
\end{table*}

The holes are therefore features formed by galactic dynamical processes (e.g.\ spurs or feathers; \citealp{kim_three-dimensional_2002,vigne_hubble_2006,dobbs_exciting_2013, WILLIAMS_PHANGSJWST}) with morphologies shaped and elongated by galactic shear, or old superbubbles created by stellar feedback that are later sheared (e.g.,\ \citealp{palous_evolution_1990}). The shells of older bubbles that have been dynamically sheared are likely the building blocks of  dust lanes, as they get compressed upon entering the arm \citep{THILKER_PHANGSJWST}. By contrast, bubbles are formed by recent stellar feedback.

To understand the link between star formation, feedback and the larger environment, this letter focuses only on bubbles currently being driven by feedback processes. While MIRI F770W provides the clearest view of holes, by itself, it does not allow us to separate the feedback-driven holes from the dynamically created holes. Combining the MIRI F770W with \HST\ imaging  preserves the high resolution view and helps delineate feedback-driven regions and dynamically created holes via the stellar populations within the shell. Furthermore, overlaying the \HST\ H$\alpha$ observations (shown as green and red in the three color image on Figure \ref{fig:bub_rgb} and Figure \ref{fig:multiwave}), we see that the spherical bubbles along the arms are usually traced by a shell of ionized gas -- whereas the elongated holes typically have little ionized gas emission. If we instead combine the higher sensitivity MUSE H$\alpha$ with F770W and \HST\ \textit{B}-band in Figure \ref{fig:bub_rgb} (\textit{B}-band highlights the young OB stars that drive bubbles), we can separate feedback-driven bubbles from dynamically driven holes.

\section{Identifying bubbles} \label{sec:method}

\noindent In this section, we outline how we identify bubbles using the multi-wavelength data sets (MIRI F770W, MUSE H$\alpha$ and HST \textit{B}-band observations) outlined in the previous section. We note here that identifying shells in F770W MIRI observations is still the primary method we use, but for small, or highly elliptical features, we rely on \HST\ sources and co-spatial MUSE H$\alpha$ to help distinguish between old and dynamically driven holes, and those more likely to be feedback-driven.

\subsection{Identification method}

\noindent To find bubble shells, two approaches are possible: automated algorithms and manual methods. Automated algorithms are reproducible and have well-defined problems and biases \citep[e.g.][]{thilker_expanding_1998, van_oort_casi_2019, silburt_lunar_2019, collischon_tracking_2021-1}. 
However they usually struggle in lower S/N situations, or when presented with complex features such as broken or elliptical shells. Furthermore, automated methods usually require fine tuning or careful training sets to output reasonable catalogs. Manual procedures are better able to utilize multi-wavelength data when identifying bubbles, and human pattern recognition is superior to automated methods at identifying weak bubble features, but they are subjective and cannot be exactly reproduced.

\JWST\ presents us with a brand new data set. Neither the exact features, nor the associated scales are known. Automatic algorithms and citizen science approaches, such as machine learning \citep{van_oort_casi_2019} and Zooniverse projects \citep{jayasinghe_milky_2019} respectively, need to be trained with already tagged catalogs. We thus opt for manual bubble identification to deliver such a tagged catalog for future works. The aim of this catalog is not, therefore, to fully characterize all the feedback-driven bubbles present, but to identify the majority of them in an as unbiased way as possible. There are undoubtedly some bubbles that we have missed, and some which are false positive detections. In Section \ref{sec:discuss}, we discuss catalog completeness further.
 
\begin{figure*}
    \centering
    \includegraphics[width = \textwidth]{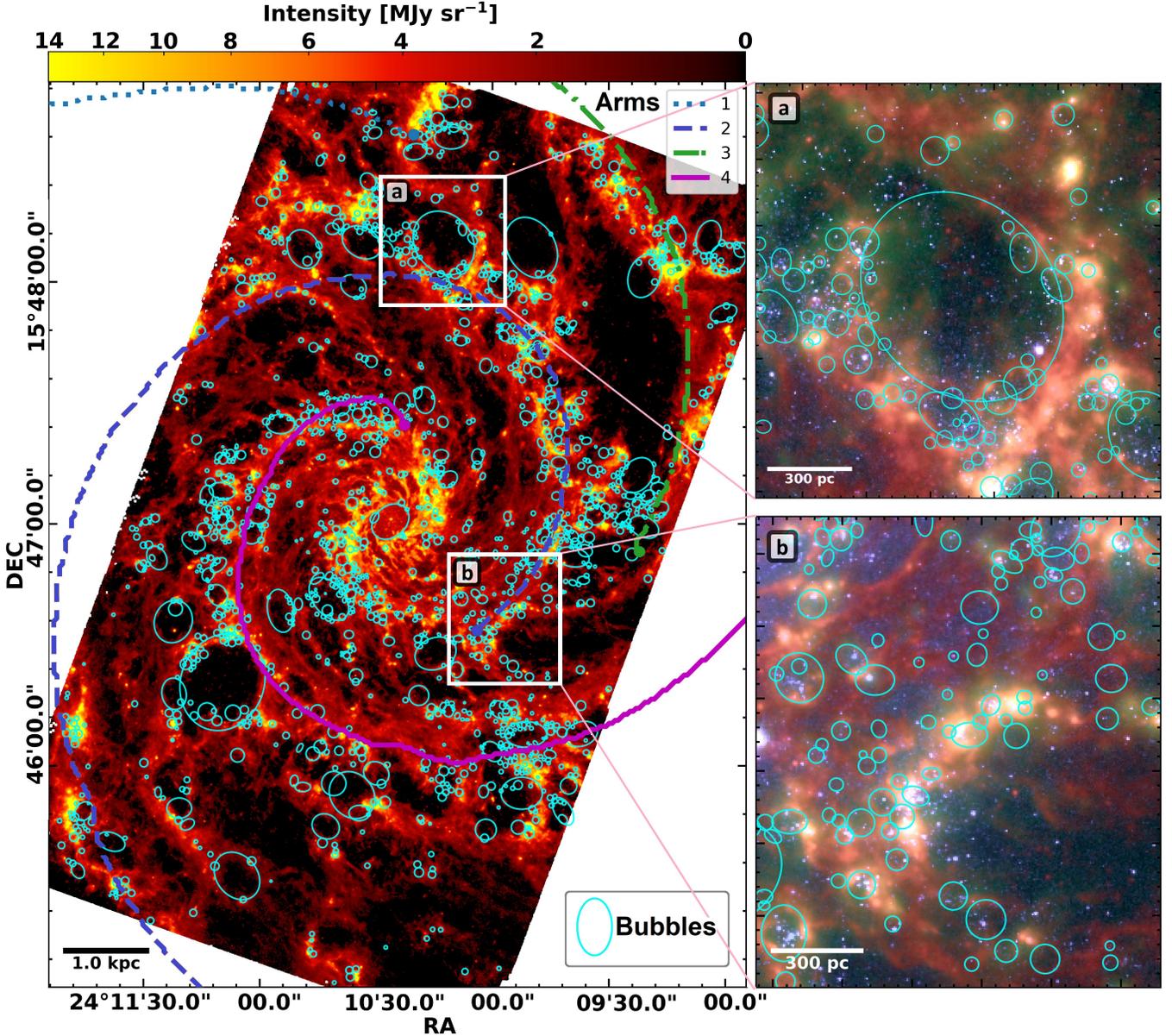}
    \caption{\textbf{Bubble locations identified in NGC~628 with multi-scale view zoom-ins providing a more detailed view}. \textbf{Left}: MIRI F770W band map of NGC~628 using a square root intensity stretch. Cyan ellipses show the location of the 1694 bubbles identified. Labeled lines indicate position of spiral arms derived by skeletonizing the arm environment masks from \cite{querejeta_stellar_2021}. Solid circle markers on arm spines indicate the start position of the arm. White boxes labeled \textit{a} and \textit{b} indicate locations of close-up panels on right side of this figure. \textbf{Right}: Same RGB as Figure \ref{fig:bub_rgb} but focused on the regions indicated on the left panel. Bars at bottom left indicates the physical size scale. }
    \label{fig:bub_loc}
\end{figure*}

We identify bubbles using an RGB image composite of MIRI F770W (red), MUSE H$\alpha$ (green), and \HST\ \textit{B}-band (blue) observations using an asinh, log, and linear stretch respectively, which we illustrate in Figure \ref{fig:bub_rgb}. We mainly focus on shell-like features present in MIRI F770W. If a feature is easy to identify as a bubble, and has a complete shell in F770W, we usually identify it without other information. For highly eccentric features, or small, clustered features, we require \HST\ sources, or H$\alpha$ emission within the bubble. For this, our definition of a shell includes any circular or elliptical feature with a radius greater than the physical resolution (6~pc), with no limit on the shell thickness in F770W. Partial features are included, such as continuous, but incomplete, circular arcs, or fragmented, clumpy circular shells. To find bubble features, we adjust the contrast for each band throughout the bubble identification process. Finally to identify bubble structures at different spatial scales, we use different zoom levels starting at a quarter of the image, down to a $\sim\ang{;;15}\times\ang{;;15}$ tile. For a more detailed discussion of how we identify bubbles, see Appendix \ref{sec:close-up}.

For each bubble, we use elliptical or circular apertures to trace the shell seen in MIRI observations (by eye), and we fit them using the shell ridge line (as opposed to the inner or outer edge of the shell). We set no limits on the eccentricity or the position angles (PAs). To speed up the bubble identification, PAs are typically incremented in 10$^\circ$ steps.

We first perform this search using one member of the team (EJW) who tried to identify most bubbles present. Two additional catalogs are generated by different team members (KH, HK) using the same method to check the robustness of our results. Section~\ref{sec:results} presents our results based on the first catalog and Appendix \ref{sec:peer-plots} lists common findings and potential variations. We note here that all  results pertaining to the bubble catalogs refer to the primary catalog (found by EJW), unless otherwise stated.

\subsection{Catalog description} \label{sec:cat-des}
\noindent The bubble catalog contains 1694 bubbles plotted on Figure~\ref{fig:bub_loc}; we tabulate the first ten in Table \ref{tab:bub_props}\footnote{All tabulated properties can be found at \url{https://www.canfar.net/storage/vault/list/phangs/RELEASES/Watkins_etal_2022}}.
The properties listed include their ID, position (in RA and Dec), their semi-major and semi-minor axes in parsec (we expect 10\% measurement uncertainties on their sizes via Watkins et al., subm), the average radii in parsec, their PA (positive angles from north to east, with $\pm\ang{5;;}$ uncertainties), which spiral arm they are closest to (as defined in Section \ref{sec:results}), the distance to this arm (where positive and negative distances are downstream and upstream respectively), and their galactocentric distance. PAs are defined between $\ang{-90;;}$ to $\ang{90;;}$. While $\ang{-90;;}$ and $\ang{90;;}$ are the same, we retain the minus sign to indicate which direction they are rounded from.
We note here that while ellipses are generally used to find bubble candidates, we use circularized average radii when summarizing their sizes in the following sections.

\section{Results} \label{sec:results}

\subsection{Global catalog properties} \label{subsec:props}

\begin{figure}
    \centering
    \includegraphics[width = 0.5\textwidth]{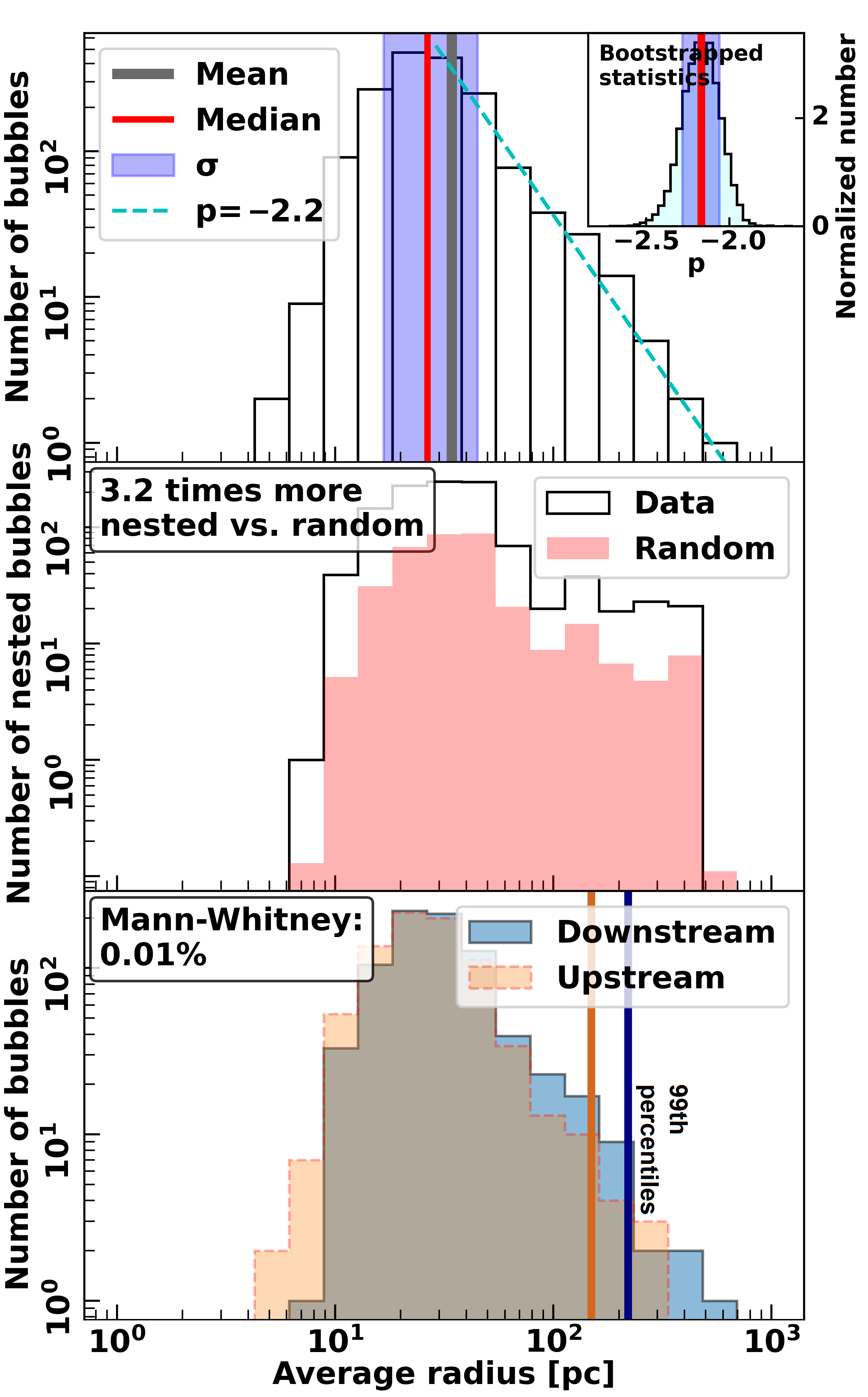}
    \caption{\textbf{Size distributions of bubbles.}  \textbf{Top}: Size distribution of the entire sample of bubbles. Solid gray and red lines are the mean and median of the distribution respectively, shaded blue region is the 16th--84th sigma range (labeled as $\sigma$ in the legend) and dashed cyan line is the power-law fit to the distribution equal to $p=-2.2$ derived using MLE. Inset shows the bootstrapped distribution of the power-law index in the bottom panel with a central value of $-2.2$ with a 16th--84th sigma range of $-2.3$ to $-2.1$. \textbf{Middle}: Total number of nested bubbles within the boundary of a larger bubble. The $x$ axis corresponds to the radius of the larger bubble. White distribution shows the data while coral shows the same distribution if instead bubble locations were randomized, averaged over 100 different randomized realizations.
    \textbf{Bottom}: Size distribution for bubbles upstream and downstream from a spiral arm shown in orange and blue respectively, with the same colored lines showing their 99th percentile values.  A Mann-Whitney test comparing the two distributions indicates 0.01\% chance that the two distributions are equal.}
    \label{fig:bub_size}
\end{figure}

\noindent Bubble radii range between 6--552~pc with a mean and median of 34 and 27~pc respectively, with a 16th--84th percentile spread of 17--45~pc. Since the average bubble diameter is 4.6--5.6 times the angular resolution of MIRI F770W, these are well resolved. The smallest bubbles also match our resolution limit, with a diameter of 12~pc, indicating that even smaller bubbles could be present for higher resolution observations.

On Figure~\ref{fig:bub_size}, we plot the size distribution of all bubbles identified. The distribution should follow a power-law of $N(R) \propto R^p$, where $N$ is the number of bubbles, $R$ are the bubble radii, and $p$ is the power-law index of the distribution. At $\sim$30~pc, the distribution visibly turns over and follows this power-law distribution. The turnover is not a physical limit to bubbles, but represents the size scale we can sample bubbles down to completely. If ambient pressures are low, some models predict the distribution can turn over at $\sim$100~pc \citep{nath_size_2020}. However, we do not expect to see such low pressures towards the center of a main sequence galaxy like NGC~628 \citep{sun_molecular_2020}, and the turnover point we measure is too low compared to the theoretical prediction.

To fit the power-law slope correctly, a precise measurement of the turnover point is needed. To find the optimal turnover point (and the corresponding $p$ value), we perform Pareto's maximum likelihood estimate (MLE) using the package \texttt{powerlaw} and re-calculate the power-law slope after removing the smallest bubble iteratively. For each solution, we perform a Kolmogorov-Smirnov (KS) test between the model and the data and choose the model that minimizes the KS value \citep{alstott_powerlaw_2014}. To quantify uncertainties on $p$ and the turnover point, we bootstrap the bubble sizes for a 1694 bubble distribution, with replacement, 10,000 times and recalculate the optimal turnover point (i.e., the minimal KS distance), and the corresponding power-law index. We find $p=-2.2\pm0.1$ with a turnover of 29$\pm$3~pc using the median bootstrapped solution (which is the same as the mean) and we quote the uncertainties using the 16th and 84th percentiles.

On smaller scales, we find bubbles are often nested within the shells of even larger bubbles with the major axis running parallel to the tangent to the shell of the larger parent bubbles. 
We quantify this nesting by counting the \textit{total} number of bubbles that intersect with the boundary of larger bubbles, binned as a function of the larger bubble's radius, and we plot this distribution in the middle panel of Figure~\ref{fig:bub_size}.
We define the bubble boundary using an annular mask 0.9--1.1 times the radius for the inner and outer edge and cross-match them with the unaltered bubble masks. If any part of the bubble mask overlaps with the annular mask of a larger bubble, we counted it as a nested bubble. However since such a comparison will result in a size bias (larger bubbles are more likely to overlap by chance due to their larger areas), we also calculate the same distribution after shuffling the positions of the bubbles 100 times and compare the two distributions in the middle panel of Figure~\ref{fig:bub_size}.

Altogether, we find 31\% of bubbles overlap with at least one smaller bubble, and bubble nesting occurs 3.2 times more often than expected from random chance, in total, when comparing the two. In future work, we will explore the exact nature of this relationship in more detail (i.e., whether the nested bubbles represent regions of new star formation, or have been simply relocated, or uncovered by the expansion of the larger bubble; see \citealt{BARNES_PHANGSJWST}).

\subsection{Azimuthal and radial trends}

\noindent When considering the locations of bubbles, Figure~\ref{fig:bub_loc} shows they are not distributed uniformly but follow the large-scale structure of the galaxy as seen by PAH emission. Specifically, bubbles are found closer to the bright emission associated with the spirals arms, with larger bubbles downstream from the arm. While an age gradient is not identified in the young stellar clusters \citep{shabani_search_2018}, we clearly see a systematic offset between star formation and molecular gas in figure 1 in \cite{kreckel_50_2018}.

To quantify what we see, we analyze the bubble properties in relation to the spiral arms. For the spiral arm model, we define a spiral arm ridge for each by skeletonizing the spiral arm environment masks calculated in \cite{querejeta_stellar_2021}, and show them on Figure~\ref{fig:bub_loc}. For each bubble outside of the galactic center (defined using the same environment masks), we identify which arm they are closest to using the difference between the bubble center and the closest approach to the spiral arm ridge. We provide these assignments in Table \ref{tab:bub_props}, keeping track of which side of the arm (upstream, downstream) they are on. With these values, we measure the separate size distributions of bubbles upstream and downstream from their nearest arm on the bottom panel of Figure~\ref{fig:bub_size}. While we find a similar number of bubbles downstream and upstream from the arms (794 vs.\ 790, respectively), the bubbles downstream are somewhat larger on average. The mean and median radii are larger by 18\% and 9\% respectively downstream, and its distribution has a more pronounced high-end tail (the radii at the 99th percentile is 220~pc for downstream bubbles, while the same for upstream bubbles is only 149~pc, a 48\% difference). 

\begin{figure}
    \centering
    \includegraphics[width = 0.5\textwidth]{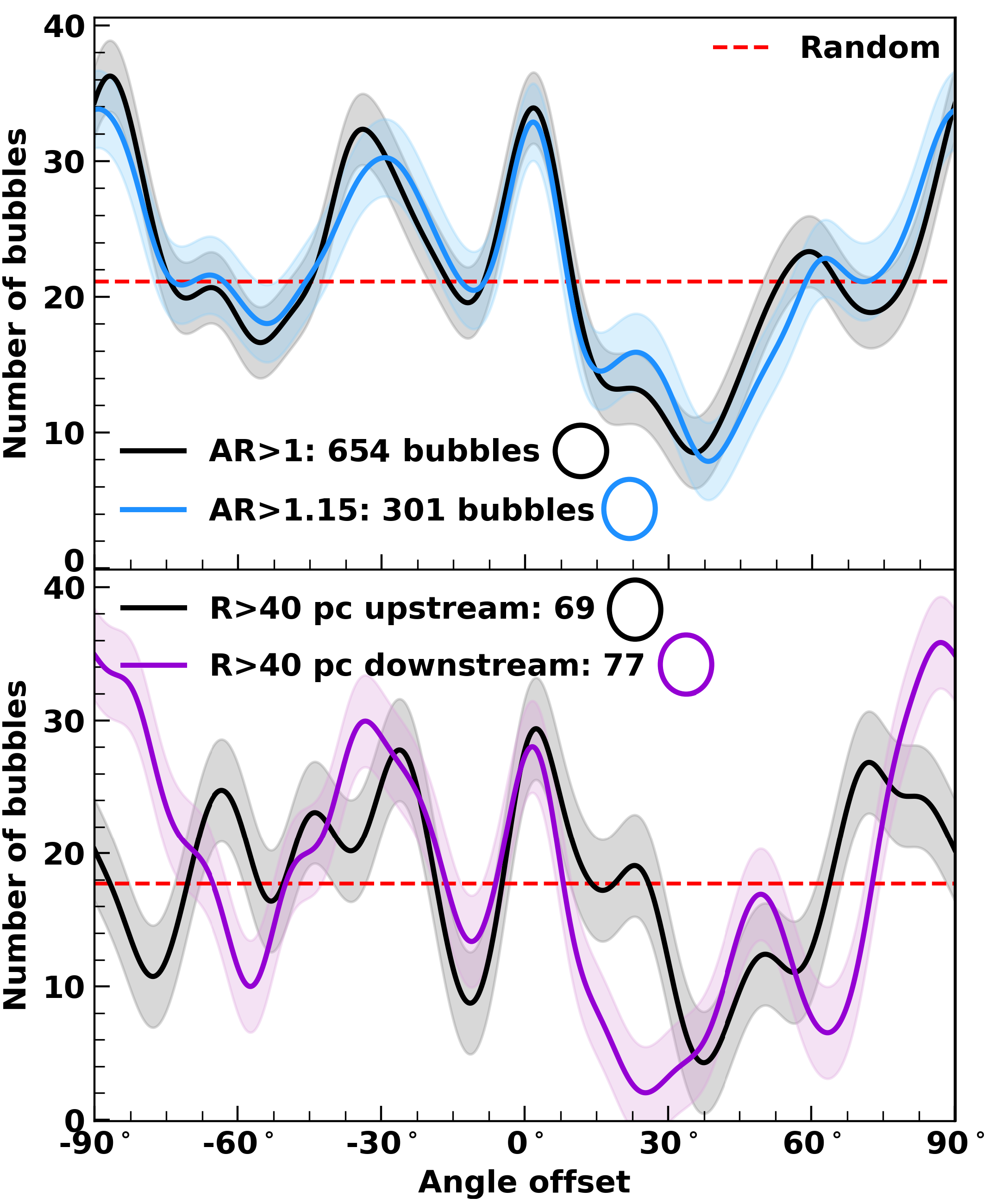}
    \caption{\textbf{Offset angles between elliptical bubbles and the spiral arm tangent angle.} \textbf{Top}: Solid black and blue lines with apertures show offset angles for bubbles with aspect ratios $>1$ and $>1.15$, respectively, for bubbles with an uncertainty in measured angle of $\pm\ang{5;;}$, resulting in a KDE-like distribution. The blue line has been re-normalized to match the area of the black line (i.e., from 301 bubbles to 654 bubbles) to make comparisons easier. Translucent regions indicate the uncertainty. Dashed red line shows the distribution of offset angles if bubbles have random PAs. \textbf{Bottom}: Same as blue line in top panel (i.e., AR$>$1.15) but for bubbles $>40$~pc upstream (black) and downstream (purple) from an arm.} 
    \label{fig:ang_off}
\end{figure}

To check how confident we can be that the two distributions are different -- informing whether there is a statistical difference between bubbles downstream from an arm compared to upstream -- we performed a Mann-Whitney test on them. The statistic indicates there is only a 0.01\% chance the two distributions have the same underlying distribution, thus we reject this null hypothesis. 
As explained in Section \ref{sec_bubsample}, we expect to see larger bubbles downstream for stars forming inside co-rotation due to the flow of gas into and through spiral arms, causing the recently formed, young stars (and the bubbles they produce) to pile up downstream. These results also reinforce that almost all the bubbles found are feedback-driven, rather than dynamical holes.

We also see that, in general, the PAs of bubbles correlate with the spiral arms. That is, the major axis of bubbles are somewhat parallel with the tangent angle of the spiral arms, though we also see some are perpendicular. If shear is responsible for non-circular bubbles, we expect PAs to correlate with the spiral arm tangent inside co-rotation \citep{palous_evolution_1990}. Bubbles offset perpendicular to the spiral arms instead represent where star formation was delayed and formed a bubble on the downstream side of the arm, which then blistered in the direction of the lower density medium downstream.

To quantify and confirm these trends, we calculate the difference ($\Delta\theta$) between the PAs of elliptical bubbles (i.e., with ARs$>1$) and the spiral arm tangent angle for bubbles outside of the galactic center and plot their distribution on the top panel of Figure~\ref{fig:ang_off} in black.  654 bubbles match these criteria. Positive $\Delta\theta$ angles are offset anti-clockwise from the arm, while clockwise $\Delta\theta$ values are negative, resulting in a $\Delta\theta$ distribution between $\ang{-90;;}$ and $\ang{90;;}$. 

We take the measurement uncertainty into consideration on Figure~\ref{fig:ang_off} by assuming the PA of each bubble has a Gaussian distribution with a standard deviation of $\ang{5;;}$ and summing their distributions together to create a kernel density estimation (KDE). We then quantify the uncertainty between the KDE and a binned histogram by calculating the mean difference between the data binned in $\ang{5;;}$ intervals and the KDE distribution generated using double the uncertainty ($\ang{10;;}$). This results in an uncertainty of $\pm$2.6 bubbles.

Figure~\ref{fig:ang_off} reveals peaks at $\ang{0;;}$ and $\ang{-30;;}$ and $\ang{-90;;}$. In addition, bubbles at $\ang{30;;}$ are under represented (and in general, positive $\Delta\theta$ are lacking). The physical meaning behind peak and trough at $\pm\ang{30;;}$ is not immediately obvious. Due to our selection criteria in Section \ref{sec:method}, the most sheared bubbles are excluded, which, if included, should have positive $\Delta\theta$ inside co-rotation as they enter the arm upstream. The peak at $\ang{-30;;}$ is likely tracing the bubbles downstream that lag behind the angle traced out by a spiral arm after it has passed.

To confirm that these peaks are statistically significant, we first estimate what Figure~\ref{fig:ang_off} would look like if PAs are randomly distributed. We randomize the bubbles PAs 100 times, recalculate $\Delta\theta$, and average thse distributions together. We find the randomization results in a uniform distribution, which we show using a dashed red line on Figure~\ref{fig:ang_off}. This test confirms that the spiral arms do not have a preferred orientation, which would result in a bias in the random distribution. Consequently, we can perform a Rayleigh test on $\Delta\theta$, which checks if a periodic distribution is non-uniform. The statistic indicates there is a 0.07\% chance that the distribution of $\Delta\theta$ is uniform, therefore we reject the null hypothesis. Finally to confirm any trends present are not driven by bubbles that might have less well defined PAs (caused by having lower ARs), we repeat our analysis after excluding bubbles with an AR $\leq1.15$ and plot the result in blue on Figure~\ref{fig:ang_off} after re-normalizing the area to improve the visual comparison. We see the same peaks and troughs, and the Rayleigh test has a p-value of 2.23\%. Altogether, these tests confirm the peaks and troughs in Figure~\ref{fig:ang_off} are real and statistically robust.

To investigate if perpendicular $\Delta\theta$ represent a population of blistering bubbles, we remake Figure~\ref{fig:ang_off} on the bottom panel for bubbles upstream and downstream. If blistering is the cause, we expect more bubbles with perpendicular $\Delta\theta$ downstream than upstream. We focus on more elliptical bubbles (with AR$>1.15$) since blistering should cause bubbles to reach higher eccentricities, and to ensure we are not dominated by smaller bubbles, we limit the analysis to bubbles $>40$~pc. Again, we re-normalize the KDEs to improve visual comparisons. Indeed, we find that bubbles downstream have perpendicular $\Delta\theta$, which are not present upstream. Moreover, the downstream distribution passes the Rayleigh test with a p-value of 2.99\%, whereas the upstream distribution fails. However in the independently generated catalogs presented in Appendix \ref{sec:peer-plots}, we do not find a strong peak at $\pm\ang{90;;}$. We expect the lower number of bubbles sampled in these catalogs is the cause. We therefore leave this as a tentative result and will explore whether bubbles align perpendicular to spiral arms in more detail in future work.

 \begin{figure}
    \centering
    \includegraphics[width = 0.5\textwidth]{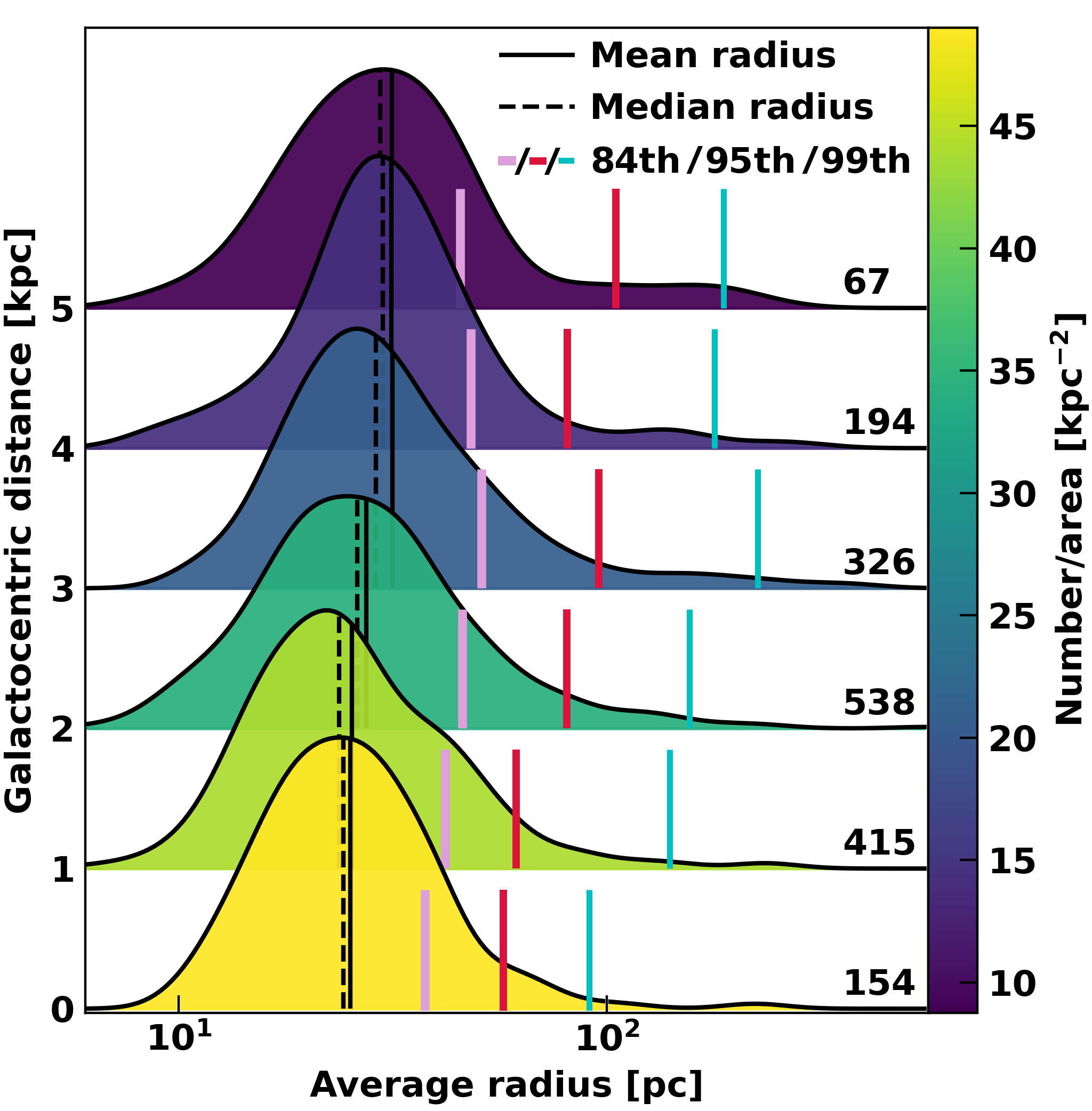}
    \caption{\textbf{Kernel Density Estimation distributions of bubble sizes binned as a function of galactocentric distance in 1~kpc rings.} Dashed and solid black lines show the median and mean values, while the solid pink, red and cyan lines shows the 84th, 95th and the 99th percentiles respectively. The color of the distributions shows the number of bubbles per unit surface area in annular rings at the distance indicated on the $y$ axis. The number labeling each distribution indicates the number of bubbles within the distribution.}
    \label{fig:rad_dist}
\end{figure}

The last statistical trend we explore is how bubble radii change as a function of galactocentric radius. Typically, we expect bubble sizes to increase further away from the center due to higher ambient pressures in galaxy centers confining the bubble sizes \citep{bagetakos_fine-scale_2011,barnes_which_2020-1}. The trend can also be driven by the orbital time. Near the center, bubbles could be sheared before they have time to grow, resulting in smaller bubbles towards the center.

Using Kendall's $\tau$ correlation coefficient, we find there is a weak correlation of 0.10 (with a p-value of $1.0\times10^{-8}$\%) between bubble radii and the galactocentric radius. However, the trend is not obvious when plotting the two against each other. To view the trend, we split the catalog into 1~kpc rings as a function of galactocentric radius and plot their distributions on Figure~\ref{fig:rad_dist}. While weak, the average radii increase as a function of galactocentric distance, and a larger fraction of bubbles are found in the high-end tail at larger galactocentric distances. Therefore, the processes that impact bubble sizes radially have a secondary role in our observations. We expect the footprint of the MIRI F770W observations impacts our ability to measure increasing bubble size with galactocentric radii. Our data only cover the inner part of the galaxy, which might impede our ability to view this trend, and the shape of the footprint also results in a poor sampling of bubbles at larger galactic radii, which we indicate by coloring the distributions on Figure \ref{fig:rad_dist}, by the number of bubbles per surface area.

\section{Discussion} \label{sec:discuss}

\subsection{Understanding what drives the shallower bubble size distribution}

\noindent In Section \ref{subsec:props}, we find the bubble size distribution is $p=-2.2\pm0.1$. Numerical simulations predict a power-law of $-2.7$ \citep{nath_size_2020} for Milky Way-like pressures, and in nearby galaxies it has been shown that this power-law can range between $-2$ to $-4$ (in H{\sc i}; \citealt{bagetakos_fine-scale_2011}). A power-law of $-2.7$ is significantly steeper than what we measure. But the size distribution of bubbles is expected to be proportional to the mechanical luminosity of OB associations \citep{oey_form_1998}, which is directly proportional to their H$\alpha$ luminosity function. For NGC~628, \cite{santoro_phangsmuse_2022} recently measured that the H{\sc ii} region H$\alpha$ luminosity function follows a power-law of $-1.7\pm0.1$. If we use this as a proxy for the OB luminosity function, it implies bubbles should follow a size distribution equal to $-2.4\pm0.2$ (see introduction in \citealt{nath_size_2020}).
However \cite{nath_size_2020} do not explore how different luminosity functions propagate in their models, so we present $-2.4\pm0.2$ as a lower limit on the theoretical value, $-2.7$.
Therefore our result differs by $1$--3.5$\sigma$, with our result leaning towards a shallower value compared to theoretical expectations (i.e., has more large bubbles and fewer small bubbles). This shows the size distribution potentially disagrees with theoretical expectations. Investigating the size distribution of more galaxies will help us confirm this conclusion.

If highly elliptical bubbles with aspect ratios $>1.5$ are instead two circular bubbles that we have misidentified, these bubbles could bias the size distribution to shallower values. To test the maximum impact this could have, we replace all bubbles with high aspect ratios ($>1.5$) with two circular bubbles with radii equal to half the semi-major axis and use MLE to find the power-law slope. We find 77 bubbles matching this criterion, and the new power-law slope is $\sim-2.3$ meaning it could account for some of the difference between observations and theory. However, this represents an extreme scenario so it is very unlikely that it explains the discrepancy we potentially observe.

Counter-intuitively, higher ambient pressures could explain the shallower slope we potentially observe (see figure 14 of \citealt{nath_size_2020}). It is a consequence of the large range of evolutionary stages that -- when averaged over time -- create the size distribution in the first place. Higher ambient pressures decrease the time it takes for bubble expansion to stall, but as a fraction of their expansion lifetime, bubbles powered by small stellar populations are more affected. Therefore the bubbles powered by larger stellar populations grow over a much longer timescale relative to the quickly stalled, smaller bubbles. As a result, at any given time, the growing bubbles occupy a larger fraction of the time averaged distribution, which makes the size distribution top-heavy \citep{nath_size_2020}. Bubble merging can also flatten the slope as it decreases the number of smaller bubbles in favor of larger bubbles.

Given we find bubbles nest together within the shells of larger bubbles, with 31\% of bubbles overlapping with at least one smaller bubble, if there is a discrepancy between theory and observations our results are consistent with bubble merging reducing the power-law slope\footnote{Although if we extrapolate the power-law index from figure 14 of \cite{nath_size_2020}, using the average dynamical pressure in NGC~628 of $\sim4.4\times10^{-12}$ dyne~cm$^{-2}$ estimated in \cite{barnes_comparing_2021} as a proxy for the ambient ISM pressure, we recover that the power-law index would be $\sim-2.1$, matching our results, though we concede there are many caveats to this estimate.}. Bubble merging is expected in galaxies, considering we see this process in the Milky Way \citep{krause_26al_2015-1} and without bubble merging, the volume bubbles occupy can exceed the total volume of galaxies when modeling their porosity \citep{clarke_galactic_2002,nath_size_2020}. Moreover, H\textsc{i} studies of nearby galaxies observe a bubble filling factor of only 10\% \citep{bagetakos_fine-scale_2011}, meaning that bubbles must merge. If we assume the difference between the theoretical value is real and caused by merging, it predicts that bubble merging reduces the power-law in NGC~628 by a minimum of $\sim$0.2, providing some constraints on the impact of bubble merging for numerically derived theories. For example the power-law index that was numerically generated in \cite{nath_size_2020} did not include the impact of bubble merging on the power-law index, even though they expect the power-law slope to become shallower if merging was included.

\subsection{Expected number of bubbles}
\noindent In the absence of merging, the number of bubbles, $N_\text{b}$ we expect to find in a galaxy is determined by three parameters: i) the star formation rate (SFR) of the galaxy; ii) the timescale, $t_\text{obs}$, over which we can \textit{observe} bubbles and; iii) the average cluster mass, $\overline{M}_*$, able to drive a bubble we can observe in this timescale (which itself depends on the form of the stellar initial mass function). Altogether we can write,

\begin{equation}
    N_\text{b} = \cfrac{\text{SFR}\times t_\text{obs}}{\overline{M}_*},
    \label{eq:num_bub}
\end{equation}

\noindent as was first outlined in \cite{clarke_galactic_2002}.

While we do not know the precise value of $t_\text{obs}$, our identification method has an implicit time limit due to bubble shearing. Bubbles that are too sheared (and are missing an obvious powering stellar population) are not identified. 
Assuming that the deviation of the bubble axis ratio ($q$, which is the inverse of the aspect ratio) from unity originates entirely from the rotation curve, we can estimate the bubble expansion velocity, $v_\text{exp}$, as a function of shear and galactocentric radius, and hence the time over which bubbles are visibly identifiable. For the exact details of the assumptions and method used in this estimate, see \cite{BARNES_PHANGSJWST} where it is presented in full.

\begin{figure}
    \centering
    \includegraphics[width = 0.5\textwidth]{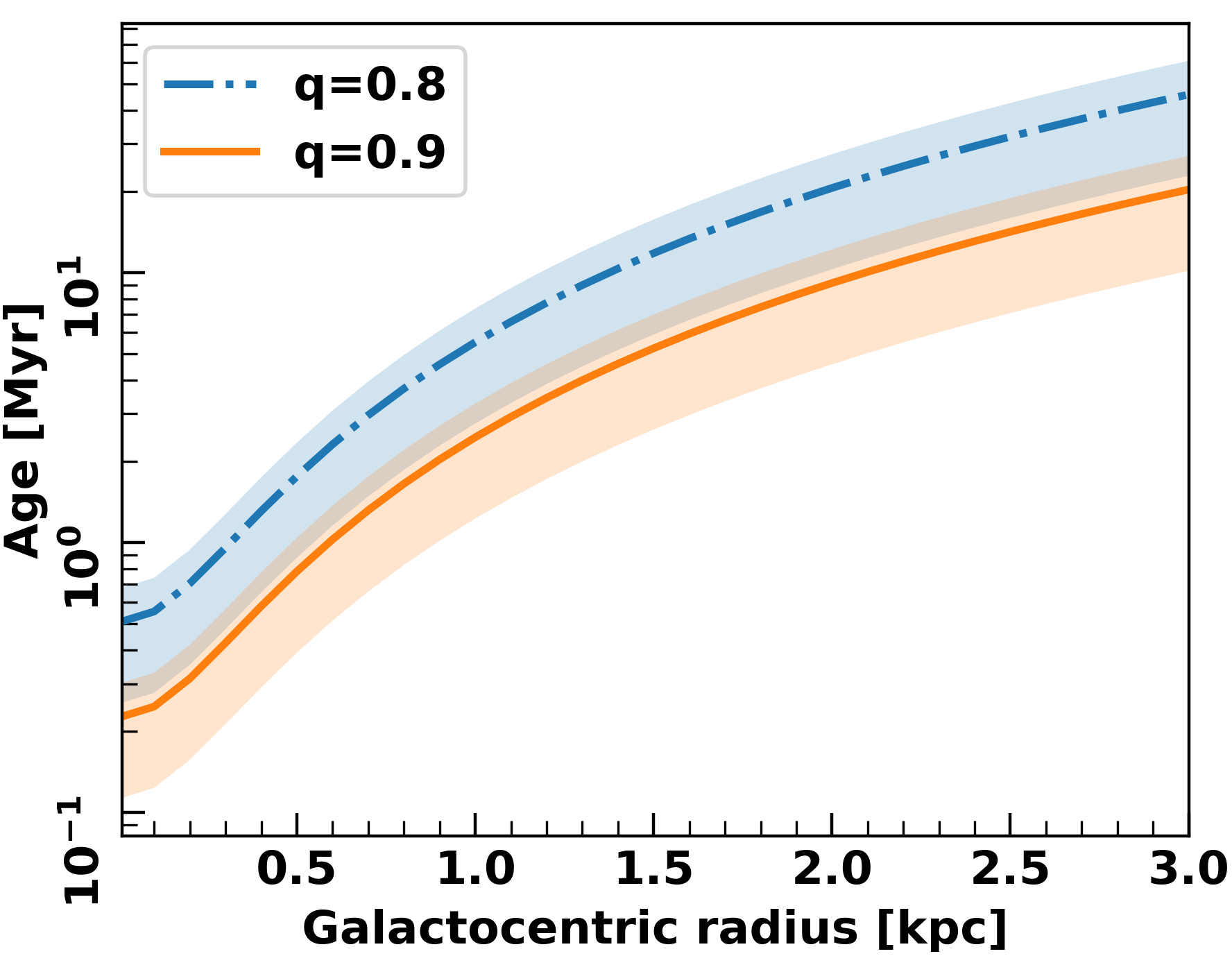}
    \caption{\textbf{Bubble age as a function of galactocentric distance}. The dashed blue and solid orange lines are the bubble ages for mean and median bubble ellipticity.}
    \label{fig:age}
\end{figure}

We find the mean and median $q$ for elliptical bubbles (i.e., bubbles identified using elliptical apertures) are 0.8 and 0.9 respectively. Assuming the bubbles are driven by feedback, we employ a self-similar expansion model to relate the resulting expansion velocities, and bubble radii, $R_\text{bub}$, to bubble age, $t_\text{age}$, using

\begin{equation}
    t_\text{age} = \eta\frac{R_\text{bub}}{v_\text{exp}},
    \label{eq:tage}
\end{equation}

\noindent and plot the resulting ages in Figure \ref{fig:age}. The constant, $\eta$, is the self-similar scaling constant \citep{ostriker_astrophysical_1988} ranging between 2/3 to 1/4 for bubbles driven purely by adiabatic winds to SN driven snowplow, respectively. Since we do not know the exact feedback mechanism driving the bubbles, we set $\eta$ to the model in the middle $\eta = 1/2$ -- which is the solution for radiative winds \citep{lancaster_efficiently_2021-1}, and use 2/3 and 1/4 as uncertainties on Figure \ref{fig:age}. We find the average bubble lies at 2.5~kpc from the galactic center, which, using the smallest and largest ages in Figure \ref{fig:age} at 2.5~kpc, predicts bubbles are visible between ages of 7--42~Myr. We note that 42~Myr is nearly identical to the expected theoretical time limit for which the largest bubbles grow. After 40~Myr, there are no OB stars left in the original stellar population that is driving the bubble.

The average cluster mass depends on the minimum cluster mass, $M_\text{min}$, that can drive bubbles above the completeness limit ($\geq$29~pc, see Section \ref{subsec:props}) in 7~Myr, and the maximum number of stars within a stellar population, $N_\text{max}$, that drives bubbles,

\begin{equation}
    \overline{M}_* = M_\text{min} \text{ln}(N_\text{max})
\end{equation}

\noindent shown in section 2.4 of \cite{clarke_galactic_2002}. From studies of nearby galaxies, the maximum number of stars we expect within a single burst has a membership of $\sim$14,000 stars (Larson et al., in subm). Using figure 9 and eq. 11 in \cite{nath_size_2020}, we predict the minimum cluster mass needed to drive a bubble  $\geq$29~pc is around $450\pm150$~\msun\ and so $\overline{M}_* = 4300\pm1400$~\msun.

Finally, the SFR of NGC~628 is 1.8~\msun~yr$^{-1}$, and considering the footprint of our MIRI observations only cover $\sim1/2$ of the total star formation \citep{leroy_phangs-alma_2021}, we use a SFR of 0.9~\msun~yr$^{-1}$ for our estimate. With observable lifetimes of 7--42~Myr, NGC~628 should contain (1000$\pm$300)--(6200$\pm$2100) bubbles $\geq$29~pc, if we expect that 30\% of bubbles that overlap merge together \citep{simpson_milky_2012,krause_26al_2015-1}. 
In our catalog we have 726 bubbles with average radii $\geq$29~pc, which is close to the lower limit of what we expect to find. While there are many assumptions that go into such a calculation, it reinforces that the features we do find at $\geq$29~pc are feedback-driven bubbles.  It also shows that a more thorough exploration of bubbles with citizen science and machine learning approaches could yield even more bubbles in NGC~628.

To quantify the observable timescale for feedback-driven bubbles we can instead ask: what timescale is needed to predict the number of bubbles present that match the number we find in a full bubble catalog (using citizen science, and machine learning approaches in future work)? Such a timescale would constrain the average energy injection into bubble shells and would constrain gas recycling times, both of which are needed to quantify mixing processes between gas and the recently enriched material provided by the SN exploding in the bubbles. 

\subsubsection{Observable bubbles in the remaining eighteen PHANGS targets}
\noindent Assuming the number of bubbles we find represents the complete number of observable bubbles in NGC~628, we can predict the total number of bubbles we should detect across the nineteen PHANGS--\JWST\ galaxies. If 787 bubbles (independently found by HK in catalog \textit{C}, see Appendix \ref{sec:peer-plots}) represents a lower bound, and 1694 an upper bound, we find the number of bubbles found per SFR is $\sim$900--1900~(\msun yr$^{-1}$)$^{-1}$. Using the SFRs given in \cite{leroy_phangs-alma_2021}, adjusted for the footprint of our \JWST\ observations\footnote{Footprints estimated using the WISE3 correction vs.\ ALMA from \cite{leroy_phangs-alma_2021}, which are approximately the same size as the \JWST\ footprints.} the total SFR across the remaining eighteen galaxies (see \citealt{LEE_PHANGSJWST} for details on the remaining eighteen targets) is 42.1~\msun~yr~$^{-1}$. Thus, we predict a further 37,000--79,000 bubbles could be found using the eighteen additional PHANGS--\JWST\ galaxies. 
NGC~628 is one of the closest and most face-on galaxies in the PHANGS--\JWST\ sample, and therefore has optimal parameters for identifying bubbles. 
Still, this represents an order of magnitude improvement on the number of bubbles we can use to investigate stellar feedback mechanisms and its connection to the gas, even when compared to bubbles identified in the Milky Way (5106 MIR bubbles found in \citealt{simpson_milky_2012}).

\section{Conclusions} \label{sec:conclude}
\noindent In this letter, we present the first high resolution (12~pc) view of feedback-driven bubbles using PAH emission in NGC~628. To find bubbles, we focus on \JWST\ MIRI F770W observations, which contains the most shell-like features compared to the other \JWST\ bands.
We combine MIRI F770W with PHANGS--\HST\ \textit{B}-band and PHANGS--MUSE H$\alpha$ observations -- tracing young stellar sources and diffuse ionized emission, both of which are signposts of recent star formation and feedback -- to help separate  feedback-driven bubbles from dynamically driven features. We find that the three data sets together help to identify the complex, hierarchical nature of the ISM, with nested bubble structures being a key characteristic of the complex multi-phase ISM.

With these three data sets, we visually identify a rich population of $\sim 1700$ bubbles with sizes spanning two orders of magnitude (6--552~pc). Locally, bubbles are highly nested, with 31\% of bubbles containing at least one smaller bubble within their shell.

We find that the size distribution follows a power-law index of $p=-2.2\pm0.1$. While within the observational expectations outlined in H\textsc{i} studies, our power-law slope might be shallower than theoretical predictions derived using the H\textsc{ii} H$\alpha$ luminosity function of NGC~628. Owing to the significant number of bubbles within the shells of larger bubbles, we conclude that, if real, bubble merging best explains why the distribution is shallower compared to theoretical predictions, which do not include the impact of bubble mergers. 

At large scales, we find azimuthal trends in the size distribution and the PAs of bubbles. Specifically, bubbles increase in size downstream from an arm compared to upstream. Since their size is related to age (i.e., their evolutionary state) bubbles are more evolved downstream, and therefore their evolution is linked to the spiral arm passage. We find that the PA are preferentially parallel or perpendicular to the spiral arms. Parallel PA are induced when expanding bubbles shear, which stretches them parallel to the arm. Potentially, the lower density gas downstream from the arm allows bubbles to blister, resulting in bubbles perpendicular to the arm though further analysis is needed to confirm this.

To characterize how well we have sampled the population, we estimate the number of bubbles we would expect to identify. We find that bubble selection is primarily limited by shear as we are usually choosing bubbles that are less elliptical. We are also limited by the resolution and find we are complete for bubbles that have radii $\geq$29~pc (where 29~pc represents our completeness limit). Given these selection criteria, we predict that we trace bubble lifetimes of 7--42~Myr and so we should find ($1000\pm300$)--($6200\pm2100$) bubbles respectively. With 726 bubbles identified with radii $\geq$29~pc, we are just within this range.

Finally, by extrapolating our result to the remaining eighteen PHANGS--\JWST\ galaxies, we predict  37,000--79,000 more bubbles could be detected. Such a number represents an order of magnitude increase in the number of resolved bubble features in the nearby universe, which will form the basis for citizen science and machine learning approaches. With comprehensive bubble catalogs, the population statistics explored in this letter can provide quantitative constraints on energy injection into the ISM as well as stellar feedback and mixing timescales.

%% Also note that the acknowledgment environment does not support long amounts of text. If you have a lot of people and institutions to acknowledge, do not use this command. Instead, create a new \section{Acknowledgments}.
%\begin{acknowledgments}
\section*{Acknowledgments}
\noindent We would like to thank the referee for their constructive
feedback that helped improve the quality of this work. This work was carried out as part of the PHANGS collaboration. Based on observations collected at the European Southern Observatory under ESO programmes 094.C-0623, 095.C-0473, and 098.C-0484.
This work also makes use of the following ALMA data: ADS/JAO.ALMA\#2012.1.00650.S, (N628/M74).
It is also been carried out on observations made with the NASA/ESA/CSA JWST. The data were obtained from the Mikulski Archive for Space Telescopes at the Space Telescope Science Institute, which is operated by the Association of Universities for Research in Astronomy, Inc., under NASA contract NAS 5-03127. The observations are associated with JWST program 2107. The data presented in this paper obtained from the Mikulski Archive for Space Telescopes (MAST) at the Space Telescope Science Institute can be accessed via: \dataset[10.17909/9bdf-jn24]{http://dx.doi.org/10.17909/9bdf-jn24}.
In addition, this research uses observations made with the NASA/ESA Hubble Space Telescope obtained from the Space Telescope Science Institute, which is operated by the Association of Universities for Research in Astronomy, Inc., under NASA contract  NAS 5–26555. These observations are associated with program 15654 and can be accessed at: \dataset[10.17909/t9-r08f-dq31]{https://dx.doi.org/10.17909/t9-r08f-dq31}.
It is also based on observations and archival data obtained with the \textit{Spitzer} Space Telescope, which is operated by the Jet Propulsion Laboratory, California Institute of Technology under a contract with NASA. All of this data used can be found at: \dataset[10.26131/IRSA414]{https://dx.doi.org/10.26131/IRSA414}.
Finally, this publication uses the data from the AstroSat mission and the UVIT instrument of the Indian Space Research Organisation (ISRO), archived at the Indian Space Science Data Centre (ISSDC). This work is supported by a grant 19ASTROSA2 from the Canadian Space Agency.
EJW acknowledges the funding provided by the Deutsche Forschungsgemeinschaft (DFG, German Research Foundation) -- Project-ID 138713538 -- SFB 881 (``The Milky Way System'', subproject P1). 
KK, OE and FS gratefully acknowledge funding from the Deutsche Forschungsgemeinschaft (DFG, German Research Foundation) in the form of an Emmy Noether Research Group (grant number KR4598/2-1, PI Kreckel). 
JMDK gratefully acknowledges funding from the European Research Council (ERC) under the European Union's Horizon 2020 research and innovation programme via the ERC Starting Grant MUSTANG (grant agreement number 714907). COOL Research DAO is a Decentralized Autonomous Organization supporting research in astrophysics aimed at uncovering our cosmic origins.
MC gratefully acknowledges funding from the DFG through an Emmy Noether Research Group (grant number CH2137/1-1).
TGW acknowledges funding from the European Research Council (ERC) under the European Union’s Horizon 2020 research and innovation programme (grant agreement No. 694343).
G.A.B. acknowledges the support from ANID Basal project FB210003.
MB acknowledges support from FONDECYT regular grant 1211000 and by the ANID BASAL project FB210003.
E.C. acknowledges support from ANID Basal projects ACE210002 and FB210003.
FB would like to acknowledge funding from the European Research Council (ERC) under the European Union’s Horizon 2020 research and innovation programme (grant agreement No.726384/Empire)
RSK, SCOG and MCS acknowledge funding from the European Research Council via the ERC Synergy Grant ``ECOGAL'' (project ID 855130), from the Deutsche Forschungsgemeinschaft (DFG) via the Collaborative Research Center ``The Milky Way System''  (SFB 881 -- funding ID 138713538 -- subprojects A1, B1, B2 and B8) and from the Heidelberg Cluster of Excellence (EXC 2181 - 390900948) ``STRUCTURES'', funded by the German Excellence Strategy. RSK also thanks the German Ministry for Economic Affairs and Climate Action for funding in project ``MAINN'' (funding ID 50OO2206). 
MQ acknowledges support from the Spanish grant PID2019-106027GA-C44, funded by MCIN/AEI/10.13039/501100011033.
ER acknowledges the support of the Natural Sciences and Engineering Research Council of Canada (NSERC), funding reference number RGPIN-2022-03499.
EWK acknowledges support from the Smithsonian Institution as a Submillimeter Array (SMA) Fellow and the Natural Sciences and Engineering Research Council of Canada.
KG is supported by the Australian Research Council through the Discovery Early Career Researcher Award (DECRA) Fellowship DE220100766 funded by the Australian Government. 
KG is supported by the Australian Research Council Centre of Excellence for All Sky Astrophysics in 3 Dimensions (ASTRO~3D), through project number CE170100013. 
AKL and NMC gratefully acknowledge support by grants 1653300 and 2205628 from the National Science Foundation, award JWST-GO-02107.009-A, award SOSP SOSPADA-010 from the NRAO, and by a Humboldt Research Award from the Alexander von Humboldt Foundation.
SKS acknowledges financial support from the German Research Foundation (DFG) via Sino-German research grant SCHI 536/11-1.

%\end{acknowledgments}

\vspace{5mm}
\facilities{HST, JWST, VLT}


\software{Python associated: Astropy (and its affiliated packages) \citep{collaboration_astropy_2013,astropy_collaboration_astropy_2018,astropy_collaboration_astropy_2022}, GeoPandas \citep{jordahl_geopandasgeopandas_2020}, scikit-image \citep{walt_scikit-image_2014}, NumPy \citep{van_der_walt_numpy_2011,harris_array_2020}, Matplotlib \citep{hunter_matplotlib_2007}, colorcet, descartes, Shapely, MultiColorFits \citep{cigan_multicolorfits_2019}, SciPy \citep{virtanen_scipy_2020}, powerlaw \citep{alstott_powerlaw_2014}. Other: SAOImage DS9 \citep{smithsonian_astrophysical_observatory_saoimage_2000,joye_new_2003}
          }



\appendix

\section{More precise rules for identifying bubbles} \label{sec:close-up}

\subsection{Bubble definition}

\noindent We begin with our definition for a bubble. These are (note, this applies to the MIRI F770W data set):
\begin{itemize}
    \item Any full or partial circular feature. Partial features can be an arc of a bubble shell, or be a clumpy but round feature;
    \item They can be any thickness, so both thinner and thicker bubbles are identified;
    \item They can be any physical size, limited only by the resolution limit of 12~pc;
    \item They can overlap and touch each other.
\end{itemize}

\begin{figure*}
    \centering
    \includegraphics[width =\textwidth]{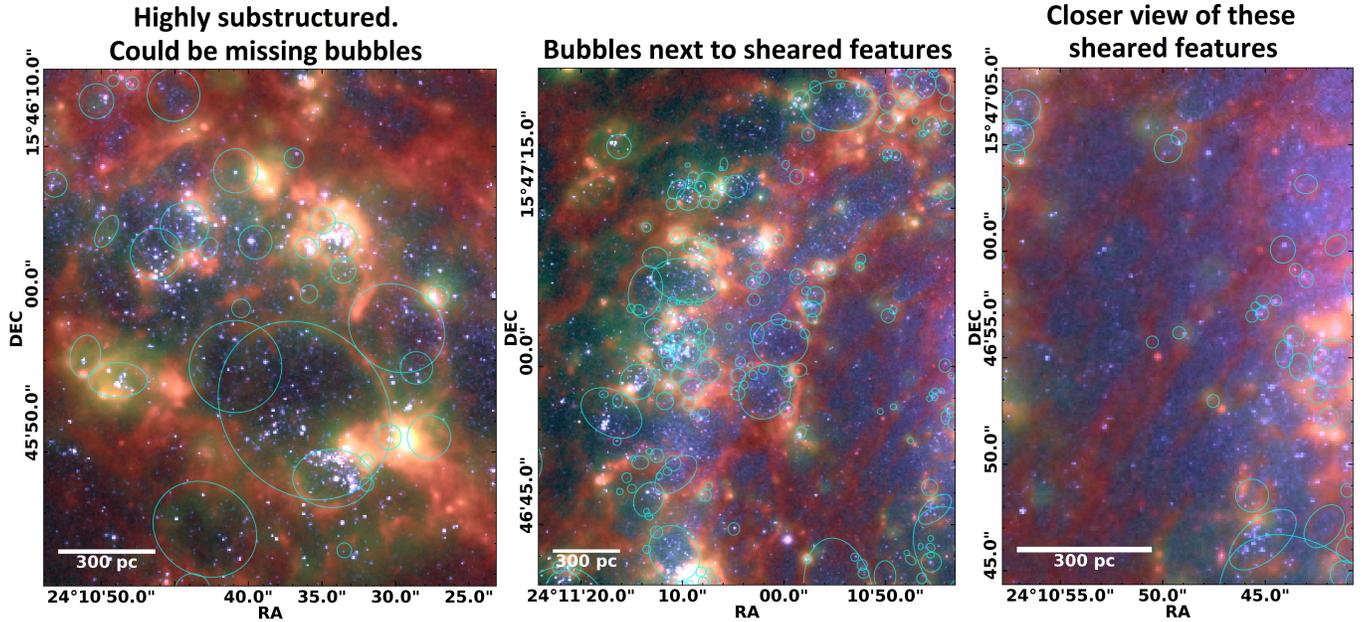}
    \caption{\textbf{Close up view of bubble features in NGC~628}. From left to right, we show: complex, nested bubbles where we might have mischaracterized the bubbles present; bubbles downstream from an arm with sheared features upstream; and a closer look at the sheared features shown on the previous panel. These RGB images have the same bands and scaling as in Figure \ref{fig:bub_rgb} and Figure \ref{fig:bub_loc}.}
    \label{fig:close-up}
\end{figure*}

\noindent If we just used these criteria, we would identify empty holes, as well as bubbles, so to differentiate between them, we loosely followed this set of rules in addition to the above:
 
\begin{itemize}
    \item We only identify stretched-out features as bubbles if they had obvious concentrated \HST\ \textit{B}-band sources within them, or required MUSE H$\alpha$ inside the hole or on the inner edge of the shell feature seen in MIRI (See Figure \ref{fig:close-up} where we show sheared features). A similar criterion was used when identifying large ($\gtrsim$300~pc) bubbles;
    \item Small partial bubble features (usually $<30$~pc) needed either co-spatial \HST\ \textit{B}-band or MUSE H$\alpha$ present within them to be identified.
\end{itemize}

Even with these rules, identifying all bubbles features is a monumental task at 12~pc resolutions due to the hierarchical nature of the ISM. Bigger bubbles were often no longer visible when we looked at them more closely (breaking up into smaller bubbles, or were not empty in emission in MIRI F770W, making them hard to define), and we often found highly nested, and complex bubble populations that were difficult to distinguish into individual bubbles (see left panel of Figure \ref{fig:close-up}). We also found that different bubbles became visible after adjusting the contrast of the MIRI F770W or the \HST\ \textit{B}-band observations, further changing what bubbles we can detect. We note this to again emphasize that when generating our catalog, the goal was not to identify every single feature that perfectly matched these criteria, but to find a \textit{representative} set of bubbles at all scales bubbles are present at. A representative sample allows us to study bubble nesting at small scales, link the location and properties of bubbles to the larger scale environment and investigate their evolution. We also note that if, for example, we added some bubbles at $\sim$70~pc scales afterwards, we would also have to search for many more smaller bubbles features to keep the catalog representative and scale unbiased.

\subsection{Detailed method for identifying bubbles}

Our identification flow followed this method:

\begin{itemize}
    \item Combine the three bands into an RGB image (red: MIRI F770W using an asinh stretch to better show diffuse shell features; green: MUSE H$\alpha$ using a log stretch since the emission spanned many orders of magnitude; blue \HST\ \textit{B}-band using a linear stretch to focus on the point-like sources);
    \item Look for bubbles, starting at a field of view (FoV) that covered 1/4 of the MIRI footprint;
    \item Fit ellipses (or circles if no strong eccentricity is visible) to the shell ridge of the bubbles present, incrementing the PA in steps of $\ang{10;;}$ to speed up the characterization process until it lined up with the shell feature. If $\ang{10;;}$ steps are insufficient, use finer increments until the ellipse traces the shell ridge;
    \item After identifying bubbles and fitting their shells, increase and decrease the contrast of the MIRI F770W band, and adjust, or add more bubbles if necessary after doing this;
    \item Blink between the RGB image with and without \HST\ \textit{B}-band to highlight any potentially missing bubbles and to check if the bubble still looked real without \HST. Remove, or add and adjust the bubbles that no longer match the definition of a bubble or whose shells now become clear respectively. This step is subjective to the individual tolerance of what someone considers a bubble to be. We note that in Watkins et al., subm, 325 superbubbles were identified both morphological and kinematically in the same set of galaxies that will be observed within PHANGS--\JWST, and so our team already has expertise in identifying bubble features in lower resolution data sets;
    \item Zoom-in and repeat (down to a tile $\sim\ang{;;15}\times\ang{;;15}$). If the larger scale bubbles no longer exist after zooming-in, but instead break up into bubble complexes, fit the smaller features and remove the larger bubble if it is no longer needed to characterize the bubble features.
\end{itemize}

\begin{figure}
    \centering
    \includegraphics[width=0.5\textwidth]{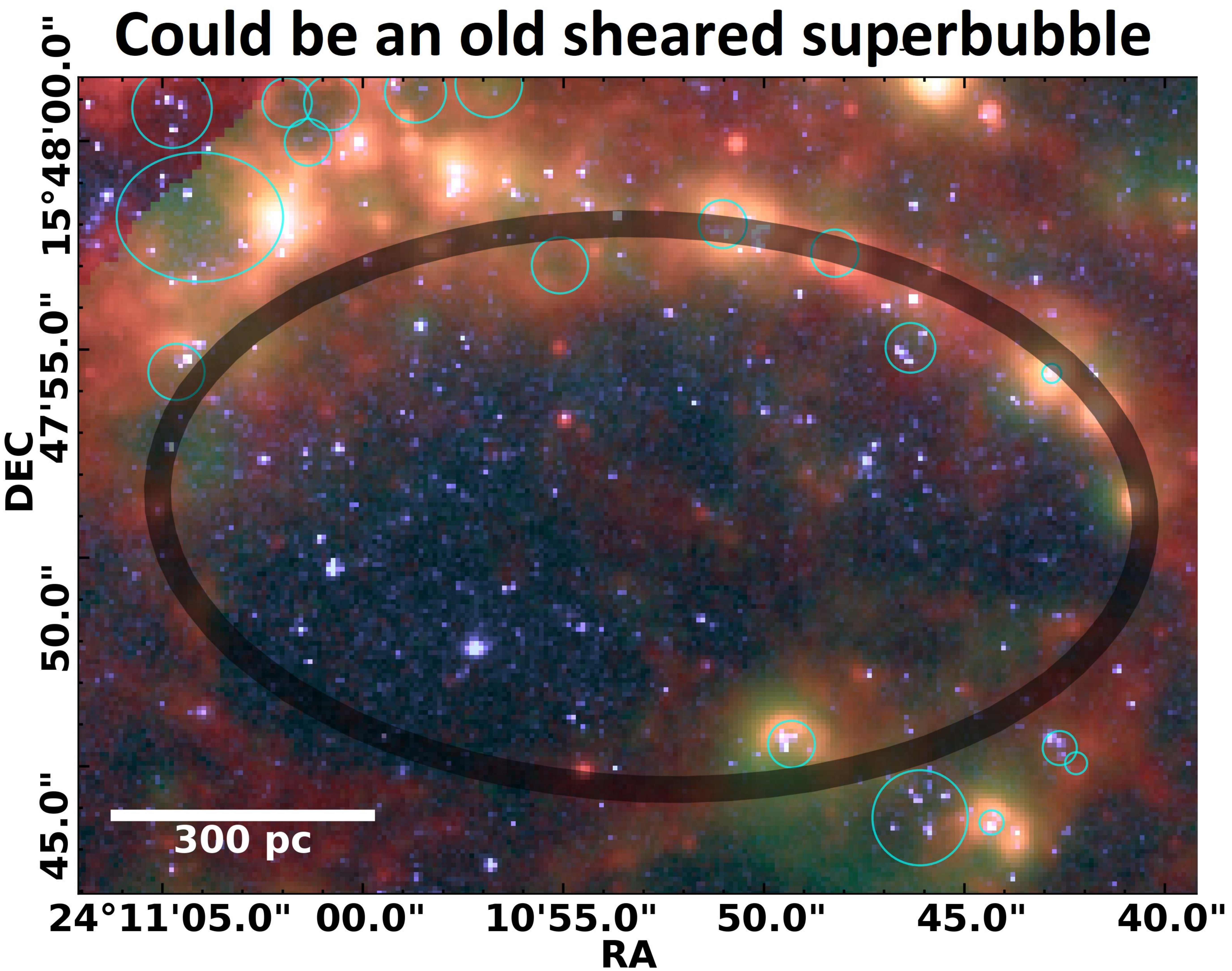}
    \caption{\textbf{Close up view of a bubble potentially missing from the catalog}. Translucent black ellipse highlights a large ($>$500~pc) feature that could be a missing large bubble that is too sheared to confirm unambiguously. RGB image with the same bands and scaling as in Figure \ref{fig:bub_rgb} and Figure \ref{fig:bub_loc}.}
    \label{fig:close-up-missing}
\end{figure}

\subsection{Difficulty in defining bubble features}

\noindent When identifying bubbles using MIRI F770W, we exclude stretched features that we are unable to confirm as bubbles using the MUSE and HST data sets. But we note that this is subjective, and there are likely features that we missed, or that are even wrongly identified as a bubble. In Section \ref{sec:discuss} we show that we are likely missing some bubbles with radii $\geq$29~pc, which we use to infer that it is unlikely that we have misidentified many bubbles $\geq$29~pc, but are instead missing bubbles.

We therefore provide more close-up examples of bubbles we identify, next to features that we deem to be too sheared to identify to help clarify our bubble finding process in Figure \ref{fig:close-up}. In the same figure, we also show where features are ambiguous and are highly nested bubbles, and in Figure \ref{fig:close-up-missing}, we show a potentially missing large ($>$500~pc) bubble. These close-up views illustrate there is no simple solution to defining bubbles, which motivates using citizen science, or machine learning techniques in the future to build up a statistically motivated catalog.

\section{Generating independent catalogs} \label{sec:peer-plots}

\begin{figure*}
     \centering
    \includegraphics[width = \textwidth]{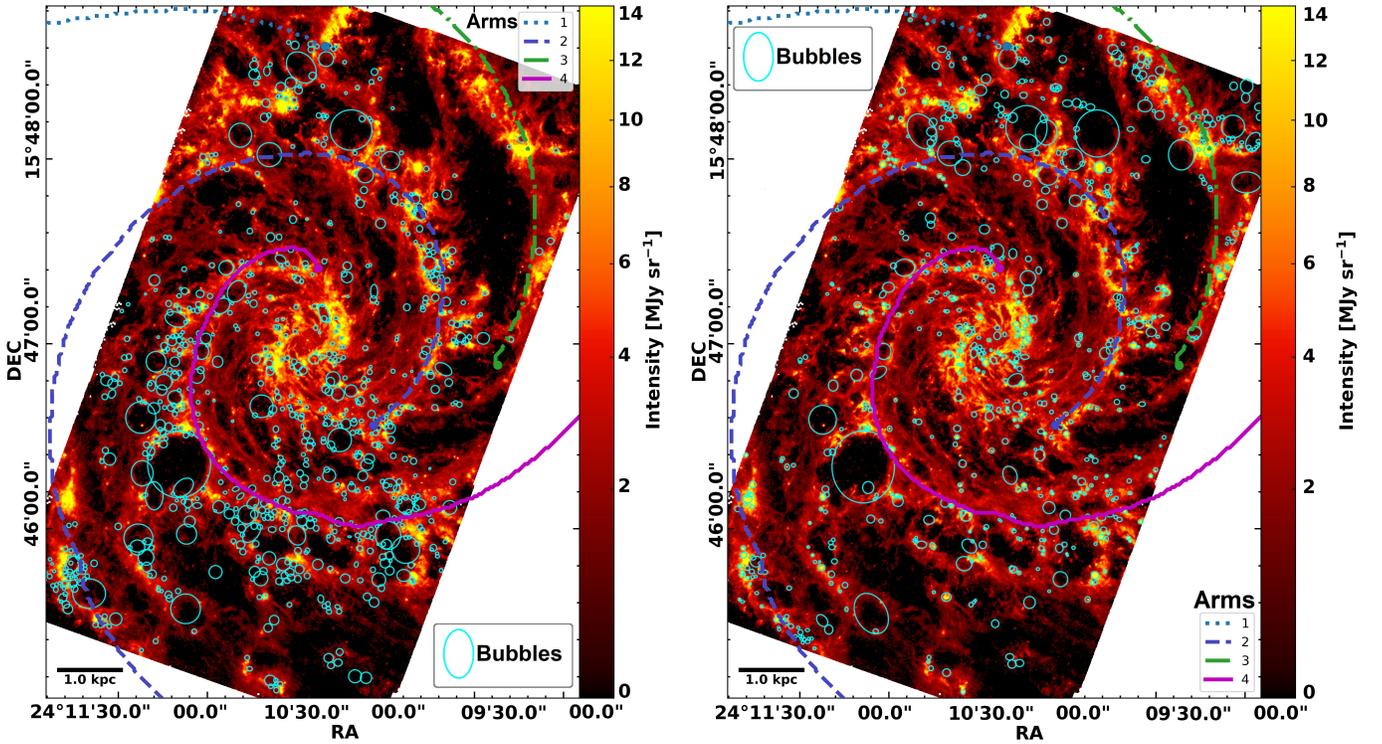}
     \caption{\textbf{Left}: Same as Figure~\ref{fig:bub_loc} for catalog \textit{B}. \textbf{Right}: Same as Figure~\ref{fig:bub_loc} for catalog \textit{C}.}
     \label{fig:ki_hw_bub_loc}
\end{figure*}

\noindent To confirm the results drawn in this letter in Section \ref{sec:results}, two additional team members independently identified bubbles, shown on Figure \ref{fig:ki_hw_bub_loc}a and Figure \ref{fig:ki_hw_bub_loc}b, following the same method outlined in Section \ref{sec:method} and Appendix \ref{sec:close-up}. These catalogs are constructed using the full map area, but with a less extensive search, 
and we use these catalogs to confirm that independent samples lead to the same conclusions (e.g., they generate the same size distribution). The purpose therefore was not to find as many bubbles as possible but to sample the bubble population present. KH and HK found 837 and 787 bubbles respectively, and we label them as catalog \textit{B} and \textit{C}, with the original catalog labeled as \textit{A} for the purpose of comparison in this section.

For catalog \textit{A}, we found 759 out of 1694 bubbles overlap (treating each bubble as a mask) with catalog \textit{B}, and from the perspective of catalog \textit{B}, 612 out of 837 bubbles overlapped with catalog \textit{A}. The difference when switching perspective reflects where a bubble overlaps with multiple bubbles. Comparing catalog \textit{A} with \textit{C}, we found 676 bubbles overlapped, and when comparing \textit{C} with \textit{A}, we found 502 out of 787 bubbles overlapped. While there is good agreement on deciding which locations have bubbles present, there are significant differences in the size, shape, or number of bubbles identified in a specific region. For instance, in some locations catalog \textit{A} characterize the structures using smaller bubble apertures than catalogs \textit{B} and \textit{C} while at the same locations catalogs \textit{B} and \textit{C} use as fewer, but larger bubble apertures. The reverse (i.e., identifying fewer, but larger bubbles in catalog \textit{A}) also occurs.

To understand if our results are robust, we repeated the analysis and figures made in the results section for the two catalogs (\textit{B} and \textit{C}). 
We find they have mean and median bubble sizes of: 35, 46; and 31, 40~pc respectively. The higher average values compared to catalog \textit{A} indicates that the peer reviewed catalogs did not identify as many small bubble features, which might impact how reliably we can compare results drawn from the small scale features.

\begin{figure*}
     \centering
    \includegraphics[width = \textwidth]{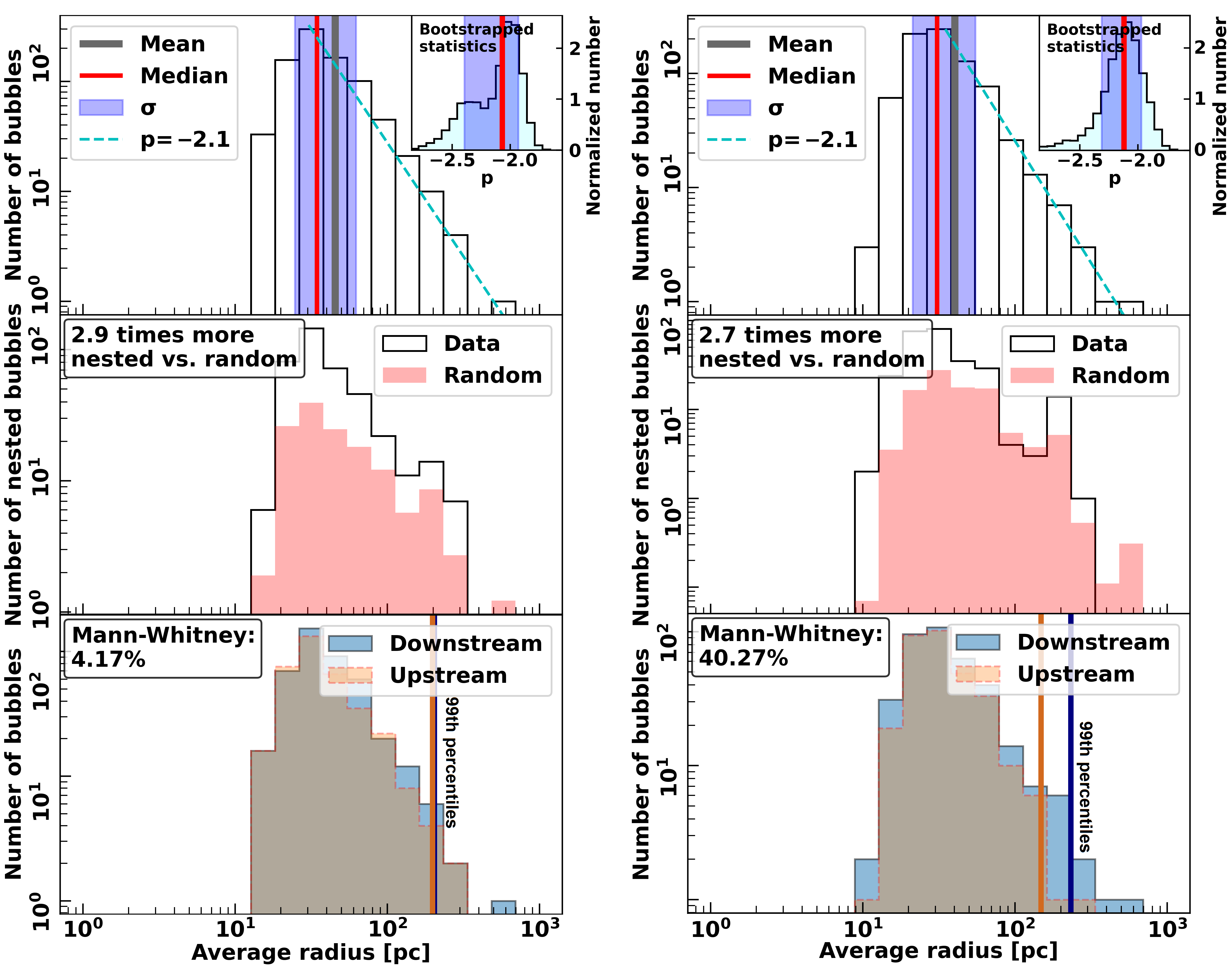}
     \caption{\textbf{Left}: Same as Figure~\ref{fig:bub_loc} for catalog \textit{B}. \textbf{Right}: Same as Figure~\ref{fig:bub_loc} for catalog \textit{C}.}
     \label{fig:ki_hw_bub_size}
\end{figure*}

We compare the overall size distributions between the three catalogs, and the power-law index derived from them in the top panels of Figure~\ref{fig:ki_hw_bub_size}a and Figure~\ref{fig:ki_hw_bub_size}b. Following the analysis in Section \ref{subsec:props}, we find all three catalogs turn over at $\sim$30~pc (\textit{B}: 30$^{+20}_{-1}$~pc, \textit{C}: 31$^{+6}_{-4}$), confirming that all three samples have a similar completeness limit. When comparing their power-laws, \textit{B} and \textit{C} have a power-law equal to $-2.1^{+0.2}_{-0.3}$ and $-2.1^{+0.1}_{-0.2}$ respectively (matching catalog \textit{A}) confirming that the power-law index we measure is robust. We note here that catalog \textit{B} has a larger upper limit for its turnover point. When inspecting the cause, we find there are two solutions for the turnover point, one at $\sim$30~pc and one at $\sim$50~pc. This indicates there is a small break in the power-law at $\sim$50~pc, with the size distribution following a slightly steeper power-law for values $>50$~pc. We do not think this is physical, but instead implies catalog \textit{B} has a size bias.

In the middle panels of Figure~\ref{fig:ki_hw_bub_size}a and Figure~\ref{fig:ki_hw_bub_size}b, we find bubble nesting is 2.9 and 2.7 times more frequent than for a random distribution in catalogs \textit{B} and \textit{C}, strongly reinforcing this result. While the number of overlaps found is smaller than in catalog \textit{A} (at 3.2), we again explain the difference by the lower number of small bubbles identified in \textit{B} and \textit{C}.

We next check the difference in bubble properties downstream and upstream from the arms in the bottom panels of Figure~\ref{fig:ki_hw_bub_size}a and Figure~\ref{fig:ki_hw_bub_size}b. Visually, the size distributions for \textit{B} and \textit{C} show a strong high-end tail and the lower number of small bubbles identified results in similar low-end tails between the upstream and downstream bubbles. As a result, the Mann-Whitney test returns a 4 and 40\% chance (for \textit{B} and \textit{C} respectively) that the two distributions sample the same underlying population. Catalog \textit{B} still allows us to reject this null hypothesis at low confidence, but at 40\%, we are unable to do the same for \textit{C}. Though, we note that both \textit{B} and \textit{C} find more bubbles in downstream (432 vs 362; 371 vs 316 downstream and upstream for \textit{B}; \textit{C} respectively), which we would expect if more bubbles are smaller upstream (given that smaller bubbles are less represented in the two catalogs). Therefore we leave our initial conclusion unchanged from Section \ref{sec:results} given that we can explain why catalogs \textit{B} and \textit{C} show a less pronounced difference.

\begin{figure*}
     \centering
    \includegraphics[width = \textwidth]{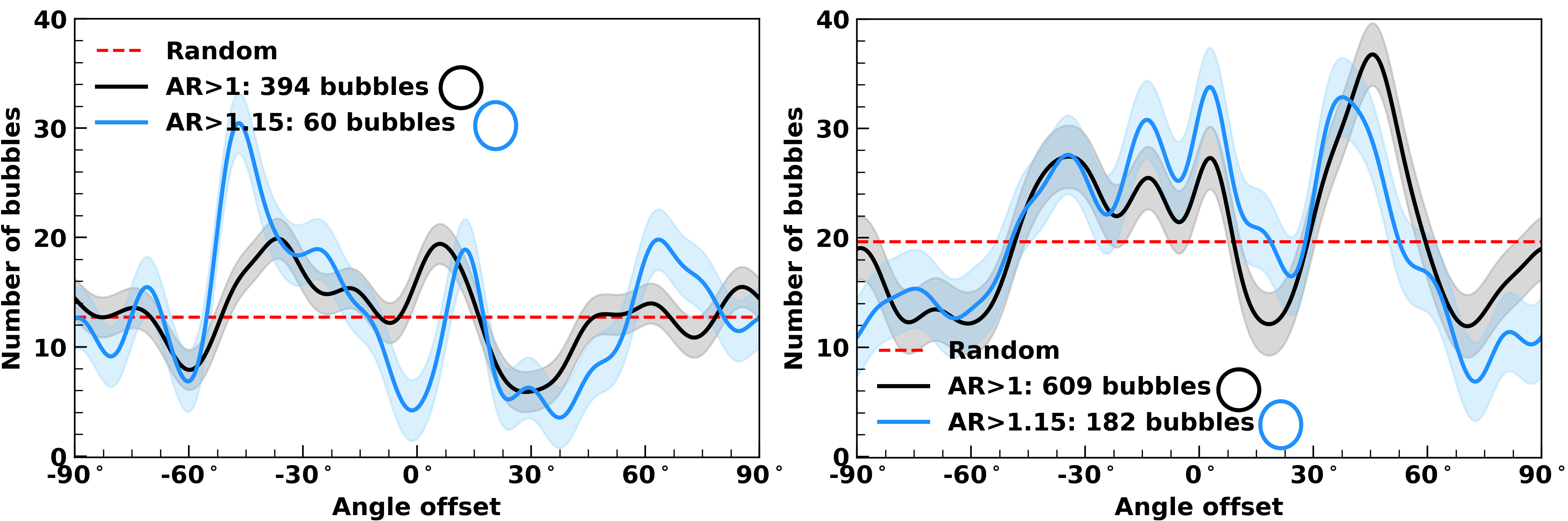}
     \caption{\textbf{Left}: Same as Figure~\ref{fig:bub_loc} for catalog \textit{B} but with the bottom panel removed. \textbf{Right}: Same as Figure~\ref{fig:bub_loc} for catalog \textit{C} but again with the bottom panel removed.}
     \label{fig:ki_hw_ang_off}
\end{figure*}

When comparing $\Delta\theta$ for catalog \textit{B} in Figure~\ref{fig:ki_hw_ang_off}a, we find similar peaks and troughs with catalog \textit{A} except catalog \textit{B} does not peak at $\pm\ang{90;;}$ and it fails the Rayleigh test, with a p-value of 6.79\%. Another difference is the number of bubbles with AR$>1.15$ in catalog \textit{B}, with 60 bubbles. In general catalog \textit{B} uses more circular apertures and lower ARs to define elliptical bubbles. The poorer sampling likely causes the missing peak at $\pm\ang{90;;}$ and the failed Rayleigh test.

Catalog \textit{C} (Figure~\ref{fig:ki_hw_ang_off}b) also differs from Catalog \textit{A}. We still find an excess of bubbles with $\Delta\theta<0$, and it also passes the Rayleigh test with a p-value of 0.09\%. However the peak at $\pm\ang{90;;}$ is not present and the trough at $\ang{30;;}$ is no longer a dominant feature. Instead we see a peak at $\ang{45;;}$. When inspecting the cause, we find the peak at $\ang{45;;}$ comes from bubbles in the outer arms (the bottom left and top right corners near Arm 2 and 3 respectively as defined in Figure \ref{fig:bub_loc}). In fact, catalog \textit{C} finds most bubbles around Arm 3, whereas the other two catalogs find the most in Arm 4 (see Figure~\ref{fig:bub_loc}, Figure~\ref{fig:ki_hw_bub_loc}a, and Figure~\ref{fig:ki_hw_bub_loc}b). If we remake Figure~\ref{fig:ki_hw_ang_off}b excluding bubbles from Arm 2 and 3, it now matches catalog \textit{B}. The outer bubbles are likely outside co-rotation defined at 4.4$\pm2.0$~kpc \citep{williams_applying_2021}, and therefore their $\Delta\theta$ might not correlate with the spiral arm passage.

In any case, the results from catalogs \textit{B} and \textit{C} inform us that we need much more complete bubble samples when analyzing trends in PAs. In future work, we will therefore undertake a more careful analysis when exploring these trends. We also note that since we do not find a peak at $\pm\ang{90;;}$, we do not compare perpendicular $\Delta\theta$ downstream vs. upstream, and so we remove this panel from Figure~\ref{fig:ki_hw_ang_off}a and Figure~\ref{fig:ki_hw_ang_off}b.

\begin{figure*}[h!]
     \centering
    \includegraphics[width = \textwidth]{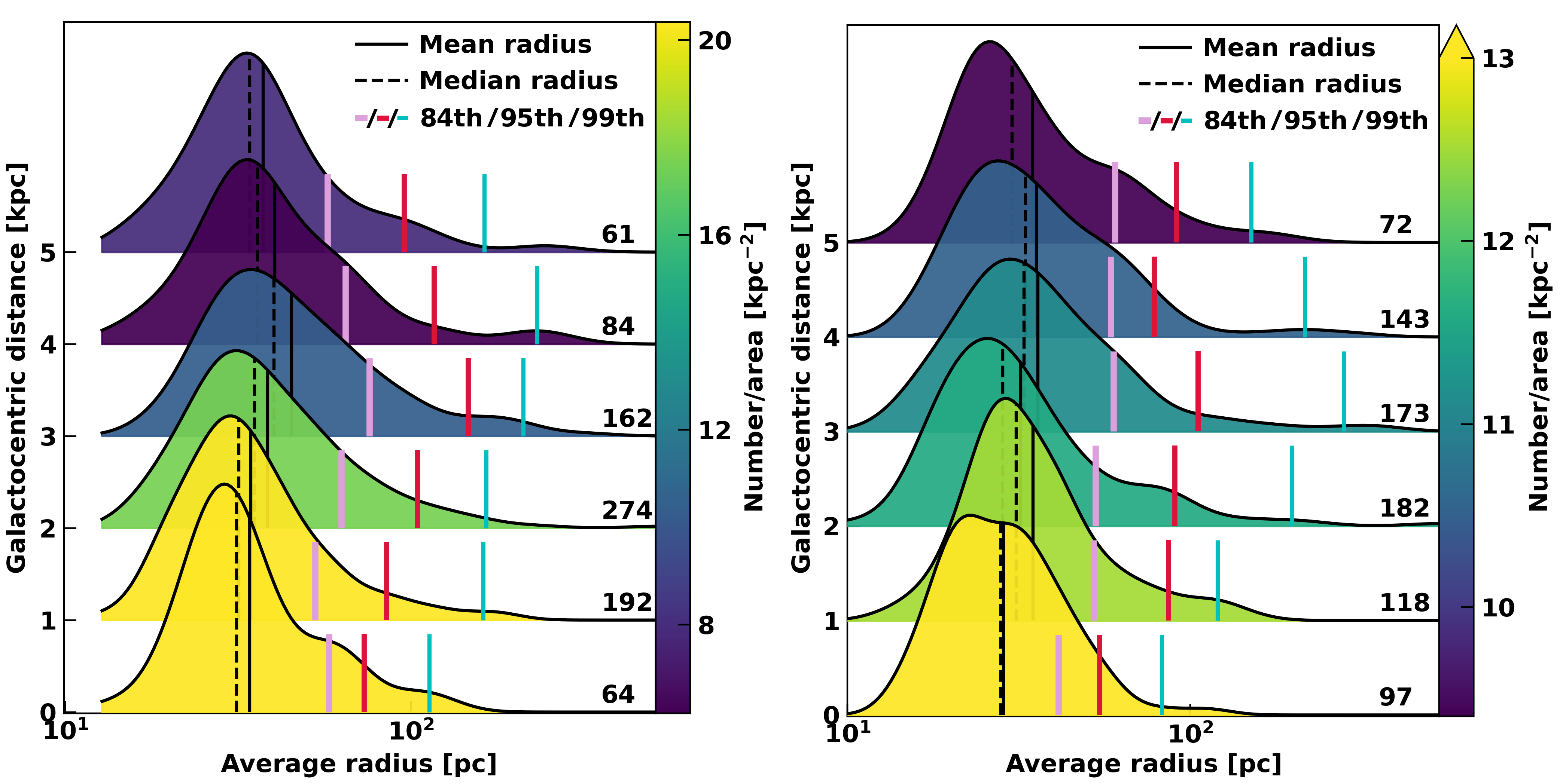}
     \caption{\textbf{Left}: Same as Figure~\ref{fig:bub_loc} for catalog \textit{B}. \textbf{Right}: Same as Figure~\ref{fig:bub_loc} for catalog \textit{C}.  We note here that the color of the distribution between 0--1~kpc is 31~kpc$^{-2}$ for catalog \textit{C}.}
     \label{fig:ki_hw_rad_dist}
\end{figure*}

Finally, we plot the radial trend with bubble size for \textit{B} and \textit{C} in Figure~\ref{fig:ki_hw_rad_dist}a and Figure~\ref{fig:ki_hw_rad_dist}b. Kendall's $\tau$ is 0.1 for both distributions, with p-values of of 0.01\% and 0.42\% for \textit{B} and \textit{C} respectively, matching \textit{A}. These indicate a weak correlation exists. We see the high-end tails of the distributions weakly increase as a function of galactocentric radius, but the number statistics at distances $>4$~pc are too poor to show a definitive increase, similar to what we see in Figure \ref{fig:rad_dist}. These results therefore reinforce the conclusions drawn previously (i.e., that the increase in bubble radius as a function of galactocentric distance is weak, indicating that the process responsible for increasing the radii is not dominant). 

%\bibliography{jwst-letter,phangsjwst}{}
\bibliography{letter}{}

\begin{thebibliography}{}
\expandafter\ifx\csname natexlab\endcsname\relax\def\natexlab#1{#1}\fi
\providecommand{\url}[1]{\href{#1}{#1}}
\providecommand{\dodoi}[1]{doi:~\href{http://doi.org/#1}{\nolinkurl{#1}}}
\providecommand{\doeprint}[1]{\href{http://ascl.net/#1}{\nolinkurl{http://ascl.net/#1}}}
\providecommand{\doarXiv}[1]{\href{https://arxiv.org/abs/#1}{\nolinkurl{https://arxiv.org/abs/#1}}}

\bibitem[{Alstott {et~al.}(2014)Alstott, Bullmore, \&
  Plenz}]{alstott_powerlaw_2014}
Alstott, J., Bullmore, E., \& Plenz, D. 2014, PLOS ONE, 9, e85777,
  \dodoi{10.1371/journal.pone.0085777}

\bibitem[{Anand {et~al.}(2021)Anand, Lee, Van~Dyk, Leroy, Rosolowsky,
  Schinnerer, Larson, Kourkchi, Kreckel, Scheuermann, Rizzi, Thilker, Tully,
  Bigiel, Blanc, Boquien, Chandar, Dale, Emsellem, Deger, Glover, Grasha,
  Groves, S.~Klessen, Kruijssen, Querejeta, {S{\'a}nchez-Bl{\'a}zquez},
  Schruba, Turner, Ubeda, Williams, \& Whitmore}]{anand_distances_2021}
Anand, G.~S., Lee, J.~C., Van~Dyk, S.~D., {et~al.} 2021, Monthly Notices of the
  Royal Astronomical Society, 501, 3621, \dodoi{10.1093/mnras/staa3668}

\bibitem[{Anderson {et~al.}(2014)Anderson, Bania, Balser, Cunningham, Wenger,
  Johnstone, \& Armentrout}]{anderson_wise_2014}
Anderson, L.~D., Bania, T.~M., Balser, D.~S., {et~al.} 2014, The Astrophysical
  Journal Supplement Series, 212, 1, \dodoi{10.1088/0067-0049/212/1/1}

\bibitem[{{Astropy Collaboration} {et~al.}(2018){Astropy Collaboration},
  {Price-Whelan}, Sip{\H o}cz, G{\"u}nther, Lim, Crawford, Conseil, Shupe,
  Craig, Dencheva, Ginsburg, VanderPlas, Bradley, {P{\'e}rez-Su{\'a}rez}, {de
  Val-Borro}, Aldcroft, Cruz, Robitaille, Tollerud, Ardelean, Babej, Bach,
  Bachetti, Bakanov, Bamford, Barentsen, Barmby, Baumbach, Berry, Biscani,
  Boquien, Bostroem, Bouma, Brammer, Bray, Breytenbach, Buddelmeijer, Burke,
  Calderone, Cano~Rodr{\'i}guez, Cara, Cardoso, Cheedella, Copin, Corrales,
  Crichton, D'Avella, Deil, Depagne, Dietrich, Donath, Droettboom, Earl, Erben,
  Fabbro, Ferreira, Finethy, Fox, Garrison, Gibbons, Goldstein, Gommers, Greco,
  Greenfield, Groener, Grollier, Hagen, Hirst, Homeier, Horton, Hosseinzadeh,
  Hu, Hunkeler, Ivezi{\'c}, Jain, Jenness, Kanarek, Kendrew, Kern, Kerzendorf,
  Khvalko, King, Kirkby, Kulkarni, Kumar, Lee, Lenz, Littlefair, Ma, Macleod,
  Mastropietro, McCully, Montagnac, Morris, Mueller, Mumford, Muna, Murphy,
  Nelson, Nguyen, Ninan, N{\"o}the, Ogaz, Oh, Parejko, Parley, Pascual, Patil,
  Patil, Plunkett, Prochaska, Rastogi, Reddy~Janga, Sabater, Sakurikar,
  Seifert, Sherbert, {Sherwood-Taylor}, Shih, Sick, Silbiger, Singanamalla,
  Singer, Sladen, Sooley, Sornarajah, Streicher, Teuben, Thomas, Tremblay,
  Turner, Terr{\'o}n, {van Kerkwijk}, {de la Vega}, Watkins, Weaver, Whitmore,
  Woillez, Zabalza, \& {Astropy
  Contributors}}]{astropy_collaboration_astropy_2018}
{Astropy Collaboration}, {Price-Whelan}, A.~M., Sip{\H o}cz, B.~M., {et~al.}
  2018, The Astronomical Journal, 156, 123, \dodoi{10.3847/1538-3881/aabc4f}

\bibitem[{{Astropy Collaboration} {et~al.}(2022){Astropy Collaboration},
  {Price-Whelan}, Lim, Earl, Starkman, Bradley, Shupe, Patil, Corrales,
  Brasseur, N{\"o}the, Donath, Tollerud, Morris, Ginsburg, Vaher, Weaver,
  Tocknell, Jamieson, {van Kerkwijk}, Robitaille, Merry, Bachetti, G{\"u}nther,
  Aldcroft, {Alvarado-Montes}, Archibald, B{\'o}di, Bapat, Barentsen,
  Baz{\'a}n, Biswas, Boquien, Burke, Cara, Cara, Conroy, Conseil, Craig, Cross,
  Cruz, D'Eugenio, Dencheva, Devillepoix, Dietrich, Eigenbrot, Erben, Ferreira,
  {Foreman-Mackey}, Fox, Freij, Garg, Geda, Glattly, Gondhalekar, Gordon,
  Grant, Greenfield, Groener, Guest, Gurovich, Handberg, Hart,
  {Hatfield-Dodds}, Homeier, Hosseinzadeh, Jenness, Jones, Joseph, Kalmbach,
  Karamehmetoglu, Ka{\l}uszy{\'n}ski, Kelley, Kern, Kerzendorf, Koch, Kulumani,
  Lee, Ly, Ma, MacBride, Maljaars, Muna, Murphy, Norman, O'Steen, Oman,
  Pacifici, Pascual, {Pascual-Granado}, Patil, Perren, Pickering, Rastogi,
  Roulston, Ryan, Rykoff, Sabater, Sakurikar, Salgado, Sanghi, Saunders,
  Savchenko, Schwardt, {Seifert-Eckert}, Shih, Jain, Shukla, Sick, Simpson,
  Singanamalla, Singer, Singhal, Sinha, Sip{\H o}cz, Spitler, Stansby,
  Streicher, {\v S}umak, Swinbank, Taranu, Tewary, Tremblay, de~{Val-Borro},
  Van~Kooten, Vasovi{\'c}, Verma, {de Miranda Cardoso}, Williams, Wilson,
  Winkel, {Wood-Vasey}, Xue, Yoachim, Zhang, Zonca, \& {Astropy Project
  Contributors}}]{astropy_collaboration_astropy_2022}
{Astropy Collaboration}, {Price-Whelan}, A.~M., Lim, P.~L., {et~al.} 2022, The
  Astrophysical Journal, 935, 167, \dodoi{10.3847/1538-4357/ac7c74}

\bibitem[{Bagetakos {et~al.}(2011)Bagetakos, Brinks, Walter, {de Blok}, Usero,
  Leroy, Rich, \& Kennicutt}]{bagetakos_fine-scale_2011}
Bagetakos, I., Brinks, E., Walter, F., {et~al.} 2011, The Astronomical Journal,
  141, 23, \dodoi{10.1088/0004-6256/141/1/23}

\bibitem[{{Barnes} {et~al.}(accepted)}]{BARNES_PHANGSJWST}
{Barnes}, A., {et~al.} accepted, \apjl

\bibitem[{Barnes {et~al.}(2020)Barnes, Longmore, Dale, Krumholz, Kruijssen, \&
  Bigiel}]{barnes_which_2020-1}
Barnes, A.~T., Longmore, S.~N., Dale, J.~E., {et~al.} 2020, Monthly Notices of
  the Royal Astronomical Society, 498, 4906, \dodoi{10.1093/mnras/staa2719}

\bibitem[{Barnes {et~al.}(2021)Barnes, Glover, Kreckel, Ostriker, Bigiel,
  Belfiore, Be{\v s}li{\'c}, Blanc, Chevance, Dale, Egorov, Eibensteiner,
  Emsellem, Grasha, Groves, Klessen, Kruijssen, Leroy, Longmore, Lopez,
  McElroy, Meidt, Murphy, Rosolowsky, Saito, Santoro, Schinnerer, Schruba, Sun,
  Watkins, \& Williams}]{barnes_comparing_2021}
Barnes, A.~T., Glover, S. C.~O., Kreckel, K., {et~al.} 2021, Monthly Notices of
  the Royal Astronomical Society, 508, 5362, \dodoi{10.1093/mnras/stab2958}

\bibitem[{Beaumont {et~al.}(2014)Beaumont, Goodman, Kendrew, Williams, \&
  Simpson}]{beaumont_milky_2014}
Beaumont, C.~N., Goodman, A.~A., Kendrew, S., Williams, J.~P., \& Simpson, R.
  2014, The Astrophysical Journal Supplement Series, 214, 3,
  \dodoi{10.1088/0067-0049/214/1/3}

\bibitem[{Belfiore {et~al.}(2022)Belfiore, Santoro, Groves, Schinnerer,
  Kreckel, Glover, Klessen, Emsellem, Blanc, Congiu, Barnes, Boquien, Chevance,
  Dale, Kruijssen, Leroy, Pan, Pessa, Schruba, \&
  Williams}]{belfiore_tale_2022}
Belfiore, F., Santoro, F., Groves, B., {et~al.} 2022, Astronomy \&
  Astrophysics, 659, A26, \dodoi{10.1051/0004-6361/202141859}

\bibitem[{Calzetti {et~al.}(2015)Calzetti, Lee, Sabbi, Adamo, Smith, Andrews,
  Ubeda, Bright, Thilker, Aloisi, Brown, Chandar, Christian, Cignoni, Clayton,
  {da Silva}, {de Mink}, Dobbs, Elmegreen, Elmegreen, Evans, Fumagalli,
  Gallagher, Gouliermis, Grebel, Herrero, Hunter, Johnson, Kennicutt, Kim,
  Krumholz, Lennon, Levay, Martin, Nair, Nota, {\"O}stlin, Pellerin, Prieto,
  Regan, Ryon, Schaerer, Schiminovich, Tosi, Van~Dyk, Walterbos, Whitmore, \&
  Wofford}]{calzetti_legacy_2015}
Calzetti, D., Lee, J.~C., Sabbi, E., {et~al.} 2015, The Astronomical Journal,
  149, 51, \dodoi{10.1088/0004-6256/149/2/51}

\bibitem[{Churchwell {et~al.}(2006)Churchwell, Povich, Allen, Taylor, Meade,
  Babler, Indebetouw, Watson, Whitney, Wolfire, Bania, Benjamin, Clemens,
  Cohen, Cyganowski, Jackson, Kobulnicky, Mathis, Mercer, Stolovy, Uzpen,
  Watson, \& Wolff}]{churchwell_bubbling_2006}
Churchwell, E., Povich, M.~S., Allen, D., {et~al.} 2006, The Astrophysical
  Journal, 649, 759, \dodoi{10.1086/507015}

\bibitem[{Cigan(2019)}]{cigan_multicolorfits_2019}
Cigan, P. 2019, Astrophysics Source Code Library, ascl:1909.002

\bibitem[{Clarke \& Oey(2002)}]{clarke_galactic_2002}
Clarke, C., \& Oey, M.~S. 2002, Monthly Notices of the Royal Astronomical
  Society, 337, 1299, \dodoi{10.1046/j.1365-8711.2002.05976.x}

\bibitem[{Collaboration {et~al.}(2013)Collaboration, Robitaille, Tollerud,
  Greenfield, Droettboom, Bray, Aldcroft, Davis, Ginsburg, {Price-Whelan},
  Kerzendorf, Conley, Crighton, Barbary, Muna, Ferguson, Grollier, Parikh,
  Nair, Unther, Deil, Woillez, Conseil, Kramer, Turner, Singer, Fox, Weaver,
  Zabalza, Edwards, Azalee~Bostroem, Burke, Casey, Crawford, Dencheva, Ely,
  Jenness, Labrie, Lim, Pierfederici, Pontzen, Ptak, Refsdal, Servillat, \&
  Streicher}]{collaboration_astropy_2013}
Collaboration, A., Robitaille, T.~P., Tollerud, E.~J., {et~al.} 2013, Astronomy
  and Astrophysics, 558, A33, \dodoi{10.1051/0004-6361/201322068}

\bibitem[{Collischon {et~al.}(2021)Collischon, Sasaki, Mecke, Points, \&
  Klatt}]{collischon_tracking_2021-1}
Collischon, C., Sasaki, M., Mecke, K., Points, S.~D., \& Klatt, M.~A. 2021,
  Astronomy \& Astrophysics, 653, A16, \dodoi{10.1051/0004-6361/202040153}

\bibitem[{Dale {et~al.}(2009)Dale, Cohen, Johnson, Schuster, Calzetti,
  Engelbracht, {Gil de Paz}, Kennicutt, Lee, Begum, Block, Dalcanton, Funes,
  Gordon, Johnson, Marble, Sakai, Skillman, {van Zee}, Walter, Weisz, Williams,
  Wu, \& Wu}]{dale_spitzer_2009}
Dale, D.~A., Cohen, S.~A., Johnson, L.~C., {et~al.} 2009, The Astrophysical
  Journal, 703, 517, \dodoi{10.1088/0004-637X/703/1/517}

\bibitem[{Dobbs \& Pringle(2013)}]{dobbs_exciting_2013}
Dobbs, C.~L., \& Pringle, J.~E. 2013, Monthly Notices of the Royal Astronomical
  Society, 432, 653, \dodoi{10.1093/mnras/stt508}

\bibitem[{Ehlerova \& Palous(1996)}]{ehlerova_origin_1996}
Ehlerova, S., \& Palous, J. 1996, Astronomy and Astrophysics, 313, 478

\bibitem[{Emsellem {et~al.}(2022)Emsellem, Schinnerer, Santoro, Belfiore,
  Pessa, McElroy, Blanc, Congiu, Groves, Ho, Kreckel, Razza,
  {Sanchez-Blazquez}, Egorov, Faesi, Klessen, Leroy, Meidt, Querejeta,
  Rosolowsky, Scheuermann, Anand, Barnes, Be{\v s}li{\'c}, Bigiel, Boquien,
  Cao, Chevance, Dale, Eibensteiner, Glover, Grasha, Henshaw, Hughes, Koch,
  Kruijssen, Lee, Liu, Pan, Pety, Saito, Sandstrom, Schruba, Sun, Thilker,
  Usero, Watkins, \& Williams}]{emsellem_phangs-muse_2022-1}
Emsellem, E., Schinnerer, E., Santoro, F., {et~al.} 2022, Astronomy \&
  Astrophysics, 659, A191, \dodoi{10.1051/0004-6361/202141727}

\bibitem[{{Faucher-Gigu{\`e}re} {et~al.}(2013){Faucher-Gigu{\`e}re}, Quataert,
  \& Hopkins}]{faucher-giguere_feedback-regulated_2013}
{Faucher-Gigu{\`e}re}, C.-A., Quataert, E., \& Hopkins, P.~F. 2013, Monthly
  Notices of the Royal Astronomical Society, 433, 1970,
  \dodoi{10.1093/mnras/stt866}

\bibitem[{Federrath(2015)}]{federrath_inefficient_2015}
Federrath, C. 2015, Monthly Notices of the Royal Astronomical Society, 450,
  4035, \dodoi{10.1093/mnras/stv941}

\bibitem[{Fielding {et~al.}(2018)Fielding, Quataert, \&
  Martizzi}]{fielding_clustered_2018}
Fielding, D., Quataert, E., \& Martizzi, D. 2018, Monthly Notices of the Royal
  Astronomical Society, 481, 3325, \dodoi{10.1093/mnras/sty2466}

\bibitem[{Fisher {et~al.}(2019)Fisher, Bolatto, White, Glazebrook, Abraham, \&
  Obreschkow}]{fisher_testing_2019}
Fisher, D.~B., Bolatto, A.~D., White, H., {et~al.} 2019, The Astrophysical
  Journal, 870, 46, \dodoi{10.3847/1538-4357/aaee8b}

\bibitem[{Grasha {et~al.}(2015)Grasha, Calzetti, Adamo, Kim, Elmegreen,
  Gouliermis, Aloisi, Bright, Christian, Cignoni, Dale, Dobbs, Elmegreen,
  Fumagalli, Gallagher, Grebel, Johnson, Lee, Messa, Smith, Ryon, Thilker,
  Ubeda, \& Wofford}]{grasha_spatial_2015}
Grasha, K., Calzetti, D., Adamo, A., {et~al.} 2015, The Astrophysical Journal,
  815, 93, \dodoi{10.1088/0004-637X/815/2/93}

\bibitem[{Grudi{\'c} {et~al.}(2019)Grudi{\'c}, Hopkins, Lee, Murray,
  {Faucher-Gigu{\`e}re}, \& Johnson}]{grudic_nature_2019}
Grudi{\'c}, M.~Y., Hopkins, P.~F., Lee, E.~J., {et~al.} 2019, Monthly Notices
  of the Royal Astronomical Society, 488, 1501, \dodoi{10.1093/mnras/stz1758}

\bibitem[{Harris {et~al.}(2020)Harris, Millman, {van der Walt}, Gommers,
  Virtanen, Cournapeau, Wieser, Taylor, Berg, Smith, Kern, Picus, Hoyer, {van
  Kerkwijk}, Brett, Haldane, {del R{\'i}o}, Wiebe, Peterson,
  {G{\'e}rard-Marchant}, Sheppard, Reddy, Weckesser, Abbasi, Gohlke, \&
  Oliphant}]{harris_array_2020}
Harris, C.~R., Millman, K.~J., {van der Walt}, S.~J., {et~al.} 2020, Nature,
  585, 357, \dodoi{10.1038/s41586-020-2649-2}

\bibitem[{Hopkins {et~al.}(2014)Hopkins, Kere{\v s}, O{\~n}orbe,
  {Faucher-Gigu{\`e}re}, Quataert, Murray, \& Bullock}]{hopkins_galaxies_2014}
Hopkins, P.~F., Kere{\v s}, D., O{\~n}orbe, J., {et~al.} 2014, Monthly Notices
  of the Royal Astronomical Society, 445, 581, \dodoi{10.1093/mnras/stu1738}

\bibitem[{Hunter(2007)}]{hunter_matplotlib_2007}
Hunter, J.~D. 2007, Computing in Science \& Engineering, 9, 90,
  \dodoi{10.1109/MCSE.2007.55}

\bibitem[{Jayasinghe {et~al.}(2019)Jayasinghe, Dixon, Povich, Binder, Velasco,
  Lepore, Xu, Offner, Kobulnicky, Anderson, Kendrew, \&
  Simpson}]{jayasinghe_milky_2019}
Jayasinghe, T., Dixon, D., Povich, M.~S., {et~al.} 2019, Monthly Notices of the
  Royal Astronomical Society, 488, 1141, \dodoi{10.1093/mnras/stz1738}

\bibitem[{Jordahl {et~al.}(2020)Jordahl, den Bossche, Fleischmann, Wasserman,
  McBride, Gerard, Tratner, Perry, Badaracco, Farmer, Hjelle, Snow, Cochran,
  Gillies, Culbertson, Bartos, Eubank, {maxalbert}, Bilogur, Rey, Ren,
  {Arribas-Bel}, Wasser, Wolf, Journois, Wilson, Greenhall, Holdgraf, Filipe,
  \& Leblanc}]{jordahl_geopandasgeopandas_2020}
Jordahl, K., den Bossche, J.~V., Fleischmann, M., {et~al.} 2020,
  Geopandas/Geopandas: V0.8.1, Zenodo, \dodoi{10.5281/zenodo.3946761}

\bibitem[{Joye \& Mandel(2003)}]{joye_new_2003}
Joye, W.~A., \& Mandel, E. 2003, 295, 489

\bibitem[{Keller {et~al.}(2022)Keller, Kruijssen, \&
  Chevance}]{keller_empirically_2022}
Keller, B.~W., Kruijssen, J. M.~D., \& Chevance, M. 2022, Monthly Notices of
  the Royal Astronomical Society, 514, 5355, \dodoi{10.1093/mnras/stac1607}

\bibitem[{{Kim} {et~al.}(subm.)}]{KIM_PHANGSJWST}
{Kim}, J., {et~al.} subm., \apjl

\bibitem[{Kim {et~al.}(2021)Kim, Ostriker, \& Filippova}]{kim_star_2021}
Kim, J.-G., Ostriker, E.~C., \& Filippova, N. 2021, The Astrophysical Journal,
  911, 128, \dodoi{10.3847/1538-4357/abe934}

\bibitem[{Kim {et~al.}(2002)Kim, Ostriker, \&
  Stone}]{kim_three-dimensional_2002}
Kim, W.-T., Ostriker, E.~C., \& Stone, J.~M. 2002, The Astrophysical Journal,
  581, 1080, \dodoi{10.1086/344367}

\bibitem[{Krause {et~al.}(2015)Krause, Diehl, Bagetakos, Brinks, Burkert,
  Gerhard, Greiner, Kretschmer, \& Siegert}]{krause_26al_2015-1}
Krause, M. G.~H., Diehl, R., Bagetakos, Y., {et~al.} 2015, Astronomy \&
  Astrophysics, 578, A113, \dodoi{10.1051/0004-6361/201525847}

\bibitem[{Kreckel {et~al.}(2018)Kreckel, Faesi, Kruijssen, Schruba, Groves,
  Leroy, Bigiel, Blanc, Chevance, Herrera, Hughes, McElroy, Pety, Querejeta,
  Rosolowsky, Schinnerer, Sun, Usero, \& Utomo}]{kreckel_50_2018}
Kreckel, K., Faesi, C., Kruijssen, J. M.~D., {et~al.} 2018, The Astrophysical
  Journal, 863, L21, \dodoi{10.3847/2041-8213/aad77d}

\bibitem[{Krumholz \& Burkhart(2016)}]{krumholz_is_2016}
Krumholz, M.~R., \& Burkhart, B. 2016, Monthly Notices of the Royal
  Astronomical Society, 458, 1671, \dodoi{10.1093/mnras/stw434}

\bibitem[{Lancaster {et~al.}(2021)Lancaster, Ostriker, Kim, \&
  Kim}]{lancaster_efficiently_2021-1}
Lancaster, L., Ostriker, E.~C., Kim, J.-G., \& Kim, C.-G. 2021, The
  Astrophysical Journal, 914, 89, \dodoi{10.3847/1538-4357/abf8ab}

\bibitem[{{Lee} {et~al.}(subm.)}]{LEE_PHANGSJWST}
{Lee}, J., {et~al.} subm., \apjl

\bibitem[{Lee {et~al.}(2022)Lee, Whitmore, Thilker, Deger, Larson, Ubeda,
  Anand, Boquien, Chandar, Dale, Emsellem, Leroy, Rosolowsky, Schinnerer,
  Schmidt, Lilly, Turner, Dyk, White, Barnes, Belfiore, Bigiel, Blanc, Cao,
  Chevance, Congiu, Egorov, Glover, Grasha, Groves, Henshaw, Hughes, Klessen,
  Koch, Kreckel, Kruijssen, Liu, Lopez, Mayker, Meidt, Murphy, Pan, Pety,
  Querejeta, Razza, Saito, {S{\'a}nchez-Bl{\'a}zquez}, Santoro, Sardone,
  Scheuermann, Schruba, Sun, Usero, Watkins, \& Williams}]{lee_phangs-hst_2022}
Lee, J.~C., Whitmore, B.~C., Thilker, D.~A., {et~al.} 2022, The Astrophysical
  Journal Supplement Series, 258, 10, \dodoi{10.3847/1538-4365/ac1fe5}

\bibitem[{{Leroy} {et~al.}(subm.)}]{LEROY1_PHANGSJWST}
{Leroy}, A., {et~al.} subm., \apjl

\bibitem[{Leroy {et~al.}(2021)Leroy, Schinnerer, Hughes, Rosolowsky, Pety,
  Schruba, Usero, Blanc, Chevance, Emsellem, Faesi, Herrera, Liu, Meidt,
  Querejeta, Saito, Sandstrom, Sun, Williams, Anand, Barnes, Behrens, Belfiore,
  Benincasa, Be{\v s}li{\'c}, Bigiel, Bolatto, {den Brok}, Cao, Chandar,
  Chastenet, Chiang, Congiu, Dale, Deger, Eibensteiner, Egorov,
  {Garc{\'i}a-Rodr{\'i}guez}, Glover, Grasha, Henshaw, Ho, Kepley, Kim,
  Klessen, Kreckel, Koch, Kruijssen, Larson, Lee, Lopez, Machado, Mayker,
  McElroy, Murphy, Ostriker, Pan, Pessa, Puschnig, Razza,
  {S{\'a}nchez-Bl{\'a}zquez}, Santoro, Sardone, Scheuermann, Sliwa, Sormani,
  Stuber, Thilker, Turner, Utomo, Watkins, \&
  Whitmore}]{leroy_phangs-alma_2021}
Leroy, A.~K., Schinnerer, E., Hughes, A., {et~al.} 2021, {{PHANGS-ALMA}}:
  {{Arcsecond CO}}(2-1) {{Imaging}} of {{Nearby Star-Forming Galaxies}}

\bibitem[{Lim {et~al.}(2020)Lim, Buizer, \& Radomski}]{lim_surveying_2020}
Lim, W., Buizer, J. M.~D., \& Radomski, J.~T. 2020, The Astrophysical Journal,
  888, 98, \dodoi{10.3847/1538-4357/ab5fd0}

\bibitem[{Marble {et~al.}(2010)Marble, Engelbracht, {van Zee}, Dale, Smith,
  Gordon, Wu, Lee, Kennicutt, Skillman, Johnson, Block, Calzetti, Cohen, Lee,
  \& Schuster}]{marble_aromatic_2010}
Marble, A.~R., Engelbracht, C.~W., {van Zee}, L., {et~al.} 2010, The
  Astrophysical Journal, 715, 506, \dodoi{10.1088/0004-637X/715/1/506}

\bibitem[{McQuinn {et~al.}(2017)McQuinn, Skillman, Dolphin, Berg, \&
  Kennicutt}]{mcquinn_accurate_2017}
McQuinn, K. B.~W., Skillman, E.~D., Dolphin, A.~E., Berg, D., \& Kennicutt, R.
  2017, The Astronomical Journal, 154, 51, \dodoi{10.3847/1538-3881/aa7aad}

\bibitem[{Nath {et~al.}(2020)Nath, Das, \& Oey}]{nath_size_2020}
Nath, B.~B., Das, P., \& Oey, M.~S. 2020, Monthly Notices of the Royal
  Astronomical Society, 493, 1034, \dodoi{10.1093/mnras/staa336}

\bibitem[{Ochsendorf {et~al.}(2015)Ochsendorf, Brown, Bally, \&
  Tielens}]{ochsendorf_nested_2015}
Ochsendorf, B.~B., Brown, A. G.~A., Bally, J., \& Tielens, A. G. G.~M. 2015,
  The Astrophysical Journal, 808, 111, \dodoi{10.1088/0004-637X/808/2/111}

\bibitem[{Oey \& Clarke(1998)}]{oey_form_1998}
Oey, M.~S., \& Clarke, C.~J. 1998, The Astronomical Journal, 115, 1543,
  \dodoi{10.1086/300290}

\bibitem[{Olivier {et~al.}(2021)Olivier, Lopez, Rosen, Nayak, Reiter, Krumholz,
  \& Bolatto}]{olivier_evolution_2021}
Olivier, G.~M., Lopez, L.~A., Rosen, A.~L., {et~al.} 2021, The Astrophysical
  Journal, 908, 68, \dodoi{10.3847/1538-4357/abd24a}

\bibitem[{Ostriker {et~al.}(2010)Ostriker, McKee, \&
  Leroy}]{ostriker_regulation_2010}
Ostriker, E.~C., McKee, C.~F., \& Leroy, A.~K. 2010, The Astrophysical Journal,
  721, 975, \dodoi{10.1088/0004-637X/721/2/975}

\bibitem[{Ostriker \& McKee(1988)}]{ostriker_astrophysical_1988}
Ostriker, J.~P., \& McKee, C.~F. 1988, Reviews of Modern Physics, 60, 1,
  \dodoi{10.1103/RevModPhys.60.1}

\bibitem[{Palous {et~al.}(1990)Palous, Franco, \&
  {Tenorio-Tagle}}]{palous_evolution_1990}
Palous, J., Franco, J., \& {Tenorio-Tagle}, G. 1990, Astronomy and
  Astrophysics, Vol. 227, p. 175-182 (1990), 227, 175

\bibitem[{Peeters {et~al.}(2004)Peeters, Spoon, \&
  Tielens}]{peeters_polycyclic_2004}
Peeters, E., Spoon, H. W.~W., \& Tielens, A. G. G.~M. 2004, The Astrophysical
  Journal, 613, 986, \dodoi{10.1086/423237}

\bibitem[{Querejeta {et~al.}(2021)Querejeta, Schinnerer, Meidt, Sun, Leroy,
  Emsellem, Klessen, {Munoz-Mateos}, Salo, Laurikainen, Beslic, Blanc,
  Chevance, Dale, Eibensteiner, Faesi, {Garcia-Rodriguez}, Glover, Grasha,
  Henshaw, Herrera, Hughes, Kreckel, Kruijssen, Liu, Murphy, Pan, Pety, Razza,
  Rosolowsky, Saito, Schruba, Usero, Watkins, \&
  Williams}]{querejeta_stellar_2021}
Querejeta, M., Schinnerer, E., Meidt, S., {et~al.} 2021, Stellar Structures,
  Molecular Gas, and Star Formation across the {{PHANGS}} Sample of Nearby
  Galaxies

\bibitem[{Reynolds \& Ogden(1979)}]{reynolds_optical_1979}
Reynolds, R.~J., \& Ogden, P.~M. 1979, The Astrophysical Journal, 229, 942,
  \dodoi{10.1086/157028}

\bibitem[{{Rodriguez} {et~al.}(accepted)}]{RODRIGUEZ_PHANGSJWST}
{Rodriguez}, J., {et~al.} accepted, \apjl

\bibitem[{S{\'a}nchez {et~al.}(2011)S{\'a}nchez, {Rosales-Ortega}, Kennicutt,
  Johnson, Diaz, Pasquali, \& Hao}]{sanchez_ppak_2011}
S{\'a}nchez, S.~F., {Rosales-Ortega}, F.~F., Kennicutt, R.~C., {et~al.} 2011,
  Monthly Notices of the Royal Astronomical Society, 410, 313,
  \dodoi{10.1111/j.1365-2966.2010.17444.x}

\bibitem[{{Sandstrom} {et~al.}(subm.)}]{SANDSTROM1_PHANGSJWST}
{Sandstrom}, K., {et~al.} subm., \apjl

\bibitem[{Santoro {et~al.}(2022)Santoro, Kreckel, Belfiore, Groves, Congiu,
  Thilker, Blanc, Schinnerer, Ho, Kruijssen, Meidt, Klessen, Schruba,
  Querejeta, Pessa, Chevance, Kim, Emsellem, McElroy, Barnes, Bigiel, Boquien,
  Dale, Glover, Grasha, Lee, Leroy, Pan, Rosolowsky, Saito, {Sanchez-Blazquez},
  Watkins, \& Williams}]{santoro_phangsmuse_2022}
Santoro, F., Kreckel, K., Belfiore, F., {et~al.} 2022, Astronomy \&
  Astrophysics, 658, A188, \dodoi{10.1051/0004-6361/202141907}

\bibitem[{Shabani {et~al.}(2018)Shabani, Grebel, Pasquali, D'Onghia, Gallagher,
  Adamo, Messa, Elmegreen, Dobbs, Gouliermis, Calzetti, Grasha, Elmegreen,
  Cignoni, Dale, Aloisi, Smith, Tosi, Thilker, Lee, Sabbi, Kim, \&
  Pellerin}]{shabani_search_2018}
Shabani, F., Grebel, E.~K., Pasquali, A., {et~al.} 2018, Monthly Notices of the
  Royal Astronomical Society, 478, 3590, \dodoi{10.1093/mnras/sty1277}

\bibitem[{Silburt {et~al.}(2019)Silburt, {Ali-Dib}, Zhu, Jackson, Valencia,
  Kissin, Tamayo, \& Menou}]{silburt_lunar_2019}
Silburt, A., {Ali-Dib}, M., Zhu, C., {et~al.} 2019, Icarus, 317, 27,
  \dodoi{10.1016/j.icarus.2018.06.022}

\bibitem[{Simpson {et~al.}(2012)Simpson, Povich, Kendrew, Lintott, Bressert,
  Arvidsson, Cyganowski, Maddison, Schawinski, Sherman, Smith, \&
  {Wolf-Chase}}]{simpson_milky_2012}
Simpson, R.~J., Povich, M.~S., Kendrew, S., {et~al.} 2012, Monthly Notices of
  the Royal Astronomical Society, 424, 2442,
  \dodoi{10.1111/j.1365-2966.2012.20770.x}

\bibitem[{Smith {et~al.}(2021)Smith, Bryan, Somerville, Hu, Teyssier, Burkhart,
  \& Hernquist}]{smith_efficient_2021}
Smith, M.~C., Bryan, G.~L., Somerville, R.~S., {et~al.} 2021, Monthly Notices
  of the Royal Astronomical Society, 506, 3882, \dodoi{10.1093/mnras/stab1896}

\bibitem[{{Smithsonian Astrophysical
  Observatory}(2000)}]{smithsonian_astrophysical_observatory_saoimage_2000}
{Smithsonian Astrophysical Observatory}. 2000, Astrophysics Source Code
  Library, ascl:0003.002

\bibitem[{Sun {et~al.}(2020)Sun, Leroy, Schinnerer, Hughes, Rosolowsky,
  Querejeta, Schruba, Liu, Saito, Herrera, Faesi, Usero, Pety, Kruijssen,
  Ostriker, Bigiel, Blanc, Bolatto, Boquien, Chevance, Dale, Deger, Emsellem,
  Glover, Grasha, Groves, Henshaw, {Jimenez-Donaire}, Kim, Klessen, Kreckel,
  Lee, Meidt, Sandstrom, Sardone, Utomo, \& Williams}]{sun_molecular_2020}
Sun, J., Leroy, A.~K., Schinnerer, E., {et~al.} 2020, The Astrophysical
  Journal, 901, L8, \dodoi{10.3847/2041-8213/abb3be}

\bibitem[{{Thilker} {et~al.}(subm.)}]{THILKER_PHANGSJWST}
{Thilker}, D., {et~al.} subm., \apjl

\bibitem[{Thilker {et~al.}(1998)Thilker, Braun, \&
  Walterbos}]{thilker_expanding_1998}
Thilker, D.~A., Braun, R., \& Walterbos, R.~M. 1998, Astronomy and
  Astrophysics, v.332, p.429-448 (1998), 332, 429

\bibitem[{Tsujimoto {et~al.}(2021)Tsujimoto, Oka, Takekawa, Iwata, Uruno,
  Yokozuka, Nakagawara, Watanabe, Kawakami, Nishiyama, Kaneko, Kanno, \&
  Ogawa}]{tsujimoto_new_2021-1}
Tsujimoto, S., Oka, T., Takekawa, S., {et~al.} 2021, The Astrophysical Journal,
  910, 61, \dodoi{10.3847/1538-4357/abe61e}

\bibitem[{{van der Walt} {et~al.}(2011){van der Walt}, Colbert, \&
  Varoquaux}]{van_der_walt_numpy_2011}
{van der Walt}, S., Colbert, S.~C., \& Varoquaux, G. 2011, Computing in Science
  \& Engineering, 13, 22, \dodoi{10.1109/MCSE.2011.37}

\bibitem[{van~der Walt {et~al.}(2014)van~der Walt, Sch{\"o}nberger,
  {Nunez-Iglesias}, Boulogne, Warner, Yager, Gouillart, \&
  Yu}]{walt_scikit-image_2014}
van~der Walt, S., Sch{\"o}nberger, J.~L., {Nunez-Iglesias}, J., {et~al.} 2014,
  PeerJ, 2, e453, \dodoi{10.7717/peerj.453}

\bibitem[{Van~Oort {et~al.}(2019)Van~Oort, Xu, Offner, \&
  Gutermuth}]{van_oort_casi_2019}
Van~Oort, C.~M., Xu, D., Offner, S. S.~R., \& Gutermuth, R.~A. 2019, The
  Astrophysical Journal, 880, 83, \dodoi{10.3847/1538-4357/ab275e}

\bibitem[{Vigne {et~al.}(2006)Vigne, Vogel, \& Ostriker}]{vigne_hubble_2006}
Vigne, M. A.~L., Vogel, S.~N., \& Ostriker, E.~C. 2006, The Astrophysical
  Journal, 650, 818, \dodoi{10.1086/506589}

\bibitem[{Virtanen {et~al.}(2020)Virtanen, Gommers, Oliphant, Haberland, Reddy,
  Cournapeau, Burovski, Peterson, Weckesser, Bright, {van der Walt}, Brett,
  Wilson, Millman, Mayorov, Nelson, Jones, Kern, Larson, Carey, Polat, Feng,
  Moore, VanderPlas, Laxalde, Perktold, Cimrman, Henriksen, Quintero, Harris,
  Archibald, Ribeiro, Pedregosa, \& {van Mulbregt}}]{virtanen_scipy_2020}
Virtanen, P., Gommers, R., Oliphant, T.~E., {et~al.} 2020, Nature Methods, 17,
  261, \dodoi{10.1038/s41592-019-0686-2}

\bibitem[{Walter {et~al.}(2008)Walter, Brinks, de~Blok, Bigiel, Kennicutt,
  Thornley, \& Leroy}]{walter_things_2008}
Walter, F., Brinks, E., de~Blok, W. J.~G., {et~al.} 2008, The Astronomical
  Journal, 136, 2563, \dodoi{10.1088/0004-6256/136/6/2563}

\bibitem[{{Williams} {et~al.}(accepted)}]{WILLIAMS_PHANGSJWST}
{Williams}, T., {et~al.} accepted, \apjl

\bibitem[{Williams {et~al.}(2021)Williams, Schinnerer, Emsellem, Meidt,
  Querejeta, Belfiore, Be{\v s}li{\'c}, Bigiel, Chevance, Dale, Glover, Grasha,
  Klessen, Kruijssen, Leroy, Pan, Pety, Pessa, Rosolowsky, Saito, Santoro,
  Schruba, Sormani, Sun, \& Watkins}]{williams_applying_2021}
Williams, T.~G., Schinnerer, E., Emsellem, E., {et~al.} 2021, The Astronomical
  Journal, 161, 185, \dodoi{10.3847/1538-3881/abe243}

\bibitem[{Zucker {et~al.}(2022)Zucker, Goodman, Alves, Bialy, Foley, Speagle,
  Gro{$\beta$}schedl, Finkbeiner, Burkert, Khimey, \&
  Swiggum}]{zucker_star_2022}
Zucker, C., Goodman, A.~A., Alves, J., {et~al.} 2022, Nature, 601, 334,
  \dodoi{10.1038/s41586-021-04286-5}

\end{thebibliography}
\bibliographystyle{aasjournal}



\suppressAffiliationsfalse
\allauthors

\end{document}